\documentclass[preprint]{aastex}
\usepackage[utf8x]{inputenc}
\usepackage{natbib}
\usepackage{graphicx}
\usepackage{subfigure}
\usepackage{textcomp}
\usepackage{amsmath,amssymb}
\usepackage[breaklinks]{hyperref}
\usepackage[all]{hypcap}
\begin{document}
\title{Determining Heating Rates in Reconnection Formed Flare Loops of the M8.0 Flare on 2005 May 13}
\author{Wen-Juan Liu\altaffilmark{1}, Jiong Qiu\altaffilmark{1}, Dana W. Longcope\altaffilmark{1}, Amir Caspi\altaffilmark{2}}
\affil{1. Department of Physics, Montana State University, Bozeman MT 59717-3840}
\affil{2. Laboratory for Atmospheric and Space Physics, University of Colorado, Boulder, CO 80303}

\begin{abstract}
We analyze and model an M8.0 flare on 2005 May 13 observed by the {\it Transition Region and Coronal Explorer} ({\it TRACE})
and {\it Reuven Ramaty High Energy Solar Spectroscopic Imager} ({\it RHESSI}) to determine the energy release rate from magnetic 
reconnection that forms and heats numerous flare loops. The flare exhibits two ribbons in UV 1600~\AA\ emission. 
Analysis shows that the UV light curve at each flaring pixel rises impulsively within a few minutes, and decays 
slowly with a timescale longer than 10~minutes. Since the lower atmosphere (the transition region and chromosphere) 
responds to energy deposit nearly instantaneously, 
the rapid UV brightening is thought to reflect the energy release 
process in the newly formed flare loop rooted at the foot point. In this paper, we utilize the spatially resolved 
(down to 1\arcsec) UV light curves and the thick-target hard X-ray emission to construct heating functions of a few 
thousand flare loops anchored at the UV foot points, and compute plasma evolution in these loops using the Enthalpy-Based 
Thermal Evolution of Loops (EBTEL) model. The modeled coronal temperatures and densities of these flare loops are then used to 
calculate coronal radiation. The computed soft X-ray spectra and light curves compare favorably with those observed 
by {\it RHESSI} and by the {\it Geostationary Operational Environmental Satellite} ({\it GOES}) X-ray Sensor (XRS). The time-dependent transition region 
differential emission measure (DEM) for each loop during its decay phase is also computed 
with a simplified model and used to calculate the optically-thin C~{\sc iv} line emission, which dominates
the UV 1600~\AA\ bandpass during the flare. The computed C~{\sc iv} line emission decays at the same
rate as observed. This study presents a method to constrain heating of reconnection-formed 
flare loops using all available observables independently, and provides insight into the physics of energy release
and plasma heating during the flare. With this method, the lower limit of the total energy 
used to heat the flare loops in this event is estimated to be 1.22
$\times$10$^{31}$~ergs, of which only 1.9
$\times$10$^{30}$~ergs is carried by beam-driven upflows during the impulsive phase, suggesting that 
the coronal plasmas are predominantly heated {\it in situ}. 

\end{abstract}
\section{Introduction}
Solar flares are generally believed to be a result of magnetic reconnection in the corona. After reconnection, the connectivity 
of the magnetic field changes and the field relaxes to a lower energy state. The energy released during reconnection is 
transported along reconnection-formed flare loops to the lower atmosphere, by conductive flux or non-thermal particles, giving
rise to enhanced emissions in optical and ultraviolet (UV) wavelengths. 
As the chromosphere is impulsively heated, a pressure front is formed that pushes plasmas both upward (chromospheric evaporation)
and downward (chromospheric condensation). The upflow fills post-flare loops, where energy is radiated away in soft X-rays (SXRs)
via thermal bremsstrahlung at temperatures of up to a few tens of million Kelvin (MK). 
When these loops cool down to a few MK, they begin to be visible in extreme ultraviolet (EUV). (See the comprehensive review by \citealt{Fletcher11}.) 

In this qualitative picture, several fundamental questions remain unanswered in terms of the quantity of energy.
First, we do not know how much energy is released by reconnection, and how much energy in total is radiated by flare plasmas 
(and carried away by coronal mass ejections, if any); in principle, the former should equal 
the latter. Second, we do not fully understand how, and by how much, flare plasmas are heated and particles are accelerated 
by reconnection-released energy, generating observed radiation signatures across the full spectral range.

A variety of hydrodynamic simulations have been performed to understand energy transport in flare loops
 and the atmospheric response to energy deposition at the feet of these loops in the transition region and chromosphere 
\citep[e.g.,\ ][]{Nagai80,Somov81,Peres82,Cheng83,Nagai84,Fisher85_6,Emslie85,Mariska89,Hawley94,Abbett99,Allred05}. 
Since plasmas are frozen in magnetic loops, most of these simulations use one-dimensional (1D) models. Some of these 
hydrodynamic simulations concentrate on the lower atmosphere by using full non-LTE formulation of radiation transfer 
\citep [i.e.,\ ][]{Fisher85_6,Hawley94,Abbett99,Allred05}, while others focus on the upper atmosphere dominated by optically 
thin emissions and treat the lower atmosphere as boundary conditions \citep [i.e.,\ ][]{Nagai80,Somov81,Peres82,Cheng83,Nagai84}. 
The radiative hydrodynamic simulations have succeeded in reproducing optically thick emissions in solar flares; however, the corona 
could only reach a few MK, which is much less than observed temperatures of tens of MK. 
On the other hand, the hydrodynamic models with optically thin emissions, though unable to address emissions from the chromosphere, 
work better at reproducing high coronal temperatures and densities, and enhancements of UV emissions from the transition region.

With high resolution EUV images obtained by the {\it Transition Region and Coronal Explorer} \citep[{\it TRACE};][]{Handy99},
it is found that the post-flare arcade is composed of at least a few hundred flaring loops formed successively \citep[e.g.,\ ][]{Aschwanden01}. 
It is further confirmed by \citet{Fletcher04} that the flare UV ribbons are made of small kernels outlining the feet of flare loops. 
To simulate sequentially formed loops in solar flares, \citet{Hori97} developed a ``pseudo-two-dimensional' model 
with multiple 1D loops heated progressively from the innermost loop to the outermost. The multi-1D model was  
further used to reproduce X-ray and EUV observations of flares \citep[e.g.,\ ][]{Hori98, Reeves02, Warren05, Warren06, Reeves07, Reeves10}. 

The hydrodynamic evolution of plasma inside the flare loop appears to be 
governed by how the loop is heated. Specifically, we need
to know when, for how long, by how much, and by what physical mechanism a flare loop is heated.
Among existing models, many start with an unspecified {\it ad hoc} heating source in the corona, and the energy is transported
through conductive flux \citep[e.g.,\ ][]{Somov82, Nagai80, Cheng83, Mariska87, Sterling93}. Some other models specify the source of
the energy and/or mechanism of heating. For example,  \citet{Somov81}, \citet{Nagai84}, \citet{Mariska89}, \citet{Emslie92}, and \citet{Reeves12} assumed that energy 
is carried by non-thermal particles and the loop is heated by Coulomb collisions of electrons with ambient plasmas, either 
in the corona (thin target) or at the lower atmosphere (thick target), using an analytic expression of electron energy 
derived by \citet{Emslie78} or \citet{Brown73}. \citet{Reeves07} and \citet{Reeves10} used Poynting flux derived from numerical 
solutions of the loss-of-equilibrium flare model \citep{Lin00,Lin04} as 
the energy input in the loop heating model, and could relate the resultant plasma radiation time profile with the kinematic 
evolution of the flux rope ejection. 

Effort has also been made to constrain the energy input with observations. \citet{Warren05} and \citet{Warren06} 
derived the heating rate empirically by matching the loop heating model results with the SXR fluxes observed by 
the X-ray Sensor (XRS) on the {\it Geostationary Operational Environmental Satellite} ({\it GOES}). In their model, the heating events are assumed to have a triangular 
time profile with a fixed duration (60~s in \citealt{Warren05}, and 400~s in \citealt{Warren06}) and their 
magnitudes and distributions during the flare are adjusted to match the observed SXR flux.
\citet{Longcope10} modeled and analyzed a flare observed by {\it TRACE}, using as the heating rates
the reconnection-released energy, which is calculated from a patchy reconnection model. The model
takes the observed photospheric magnetic field as the boundary, and uses the observationally measured
reconnection rate and time distribution of reconnection-formed flare loops counted in EUV images
obtained by {\it TRACE}. The modeled coronal radiation compares favorably with SXR observations from {\it GOES} XRS and the {\it Reuven Ramaty High Energy Solar Spectroscopic Imager} \citep[{\it RHESSI};][]{Lin02}.

Along this avenue, recently, \citet{Qiu12} have proposed that the heating rates in individual flare loops 
could be directly inferred from the time profiles of UV emission at the feet of flare loops. 
The spatially resolved UV light curves exhibit an impulsive rise, which indicates, and therefore constrains, 
when, for how long, and by how much a flare loop is heated. Accordingly, \citet{Qiu12} constructed 
from these light curves the heating rates of 1600 loops, {\it each} with cross-sectional area of 1\arcsec\ by 1\arcsec, anchored at 
impulsively brightened UV pixels in a C3.2 flare observed by the Atmospheric Imaging Assembly 
\citep[AIA;][]{Lemen12} onboard the {\it Solar Dynamic Observatory} \citep[{\it SDO};][]{Pesnell12}. Applying these heating rates to
a zero-dimensional loop heating model, the Enthalpy-Based Thermal Evolution of Loops 
\citep[EBTEL;][]{Klimchuk08, Cargill2012_2, Cargill2012_3} model,
they computed plasma evolution in these 1600 flare loops and the synthetic coronal radiation
in SXR and EUV passbands, which are compared with observations by {\it GOES} and AIA to verify the empirically
determined heating rates. This method uses all available observations to constrain the heating rates from the input
(the impulsive UV emission from the foot points) to the output (the X-ray and EUV emission from the coronal loops). The flare studied in their paper is
primarily a thermal flare with little hard X-ray (HXR) emission, and {\it ad hoc} volumetric heating rates are used in the model.

In this paper, we will improve the method of \citet{Qiu12} and apply the analysis to an M8.0 flare
on 2005 May 13 observed by {\it TRACE}, {\it GOES}, and {\it RHESSI}. 
The flare exhibits significant thick-target non-thermal HXR emission observed by {\it RHESSI}, 
suggestive of strong direct heating of the lower atmosphere that would result in chromospheric evaporation 
that sends energy back to the corona. We also include in the loop heating model this 
energy flux of non-thermal origin, using observed UV and thick-target HXR emissions as constraints. 
As the output of the model, we calculate the time-dependent  3--20~keV SXR spectrum and compare this with the observations 
by {\it RHESSI}. Finally, whereas the impulsive rise of the UV foot-point emission is considered 
to directly relate to the energy release process, the observed gradual decay of the UV emission from the same 
foot-point is governed by evolution of the overlying coronal plasma in the loop, which therefore provides 
diagnostics of the loop cooling. In this paper, we use a simplified model to compute the UV emission 
during the decay of flare loops to further compare with observations. These new steps help improve determination 
of the heating rates, and enhance our understanding of flare energy release, partition, and plasma heating and evolution.
In the following text, we present in Section~\ref{Sec2} observations of the flare, especially the UV and HXR observations, which will be used to 
construct the heating functions, and apply them to model coronal plasma evolution in Section~\ref{Sec3}. In Section~\ref{Sec4}, 
we compute time-dependent SXR and UV fluxes from the model output and compare them with observations. 
Conclusions and discussions are given in Section~\ref{Sec5}.

\section{Observations and Analysis}\label{Sec2}

\subsection{Overview of Observations}\label{Sec2.1}
\begin{figure}
\plotone{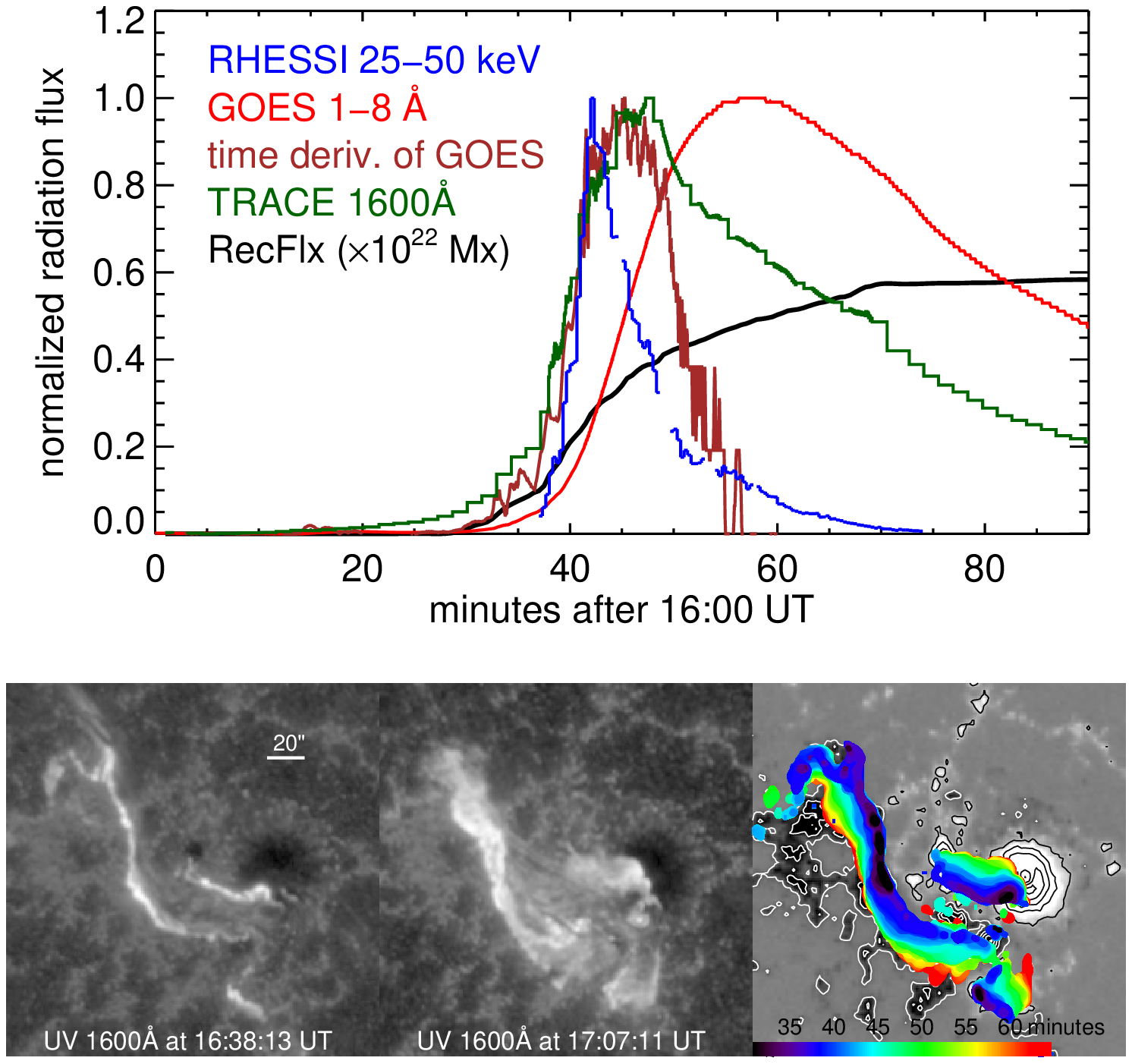}
\caption{Light curves and images of the M8.0 flare observed on 2005 May 13. The upper panel shows background-subtracted and normalized
light curves in HXRs (25--50~keV) observed by {\it RHESSI} detector 4 (blue), SXRs (1--8~\AA{}) by {\it GOES} XRS (red),
and UV (1600~\AA{} band) by {\it TRACE} (dark green). Also plotted are the time derivatives of the 1--8~\AA\ SXR flux (brown), and magnetic
reconnection flux (black) measured by using UV 1600~\AA{} observations and the longitudinal magnetogram from {\it SOHO}/MDI. 
The left and middle images in the lower panel are snapshots of UV images during the impulsive and decay phases of
the flare observed by {\it TRACE}, while the lower right panel shows evolution of the UV brightening on top of the MDI longitudinal 
magnetogram taken at 16:03:02~UT with the color bar indicating the start times of UV brightening at different locations.
} \label{Fig1}
\end{figure}

In this paper, we study a {\it GOES}-class M8.0 flare that occurred on 2005 May 13 in NOAA Active Region 10759 
located at N12E05 at the time of the flare. Figure~\ref{Fig1} summarizes observations of the flare
in a few wavelengths. High cadence (3~s) UV images obtained by {\it TRACE}, given in the bottom panel, reveal that 
the M8.0 flare is a typical two ribbon flare. The flare UV ribbons expand away from the magnetic 
polarity inversion line for half an hour, indicating that magnetic reconnection continues to form
new loops and release energy in them, as depicted by the standard flare model 
\citep[the CSHKP model;][]{Carmichael64, Sturrock66, Hirayama74, Kopp76}. 
Significant X-ray emissions are observed in this flare by {\it RHESSI}. Previous morphology studies of this same 
flare show that, during the impulsive phase, the sources of non-thermal HXR emissions of $\ge$25 keV 
are located within the UV flare ribbons \citep[see Fig.~2 and Fig.~3 of][for images of 
HXR kernels in 25--50~keV and 50--100~keV energy bands overlaid on the UV contours]{Liu07apjl}. 
These observations confirm that UV and HXR emissions are both produced 
by heating of the lower atmosphere (the transition region and chromosphere) by energetic particles precipitating 
at the feet of flare loops during the impulsive phase \citep[see, e.g.,\ ][]{Cheng81, Cheng88, Warren2001, Coyner09, Cheng12}. 

The top panel in Figure~\ref{Fig1} shows normalized and background-subtracted X-ray and UV light 
curves observed by {\it GOES}, {\it RHESSI}, and {\it TRACE}, respectively. The UV light curve is the total count 
rate (in units of DN s$^{-1}$) derived from the semi-calibrated UV images 
\citep[see][Section~3.1 for calibration techniques]{Qiu10}, with pre-flare counts subtracted. 
It follows closely the $\ge$25~keV HXR light curve
during the rise of the flare and the impulsive phase, but continues to rise after the HXR peak at 16:42~UT, and 
reaches maximum five minutes later. Also plotted is the time derivative of the 
{\it GOES} 1--8~\AA\ SXR light curve, and it is seen that the UV light curve peaks at 
the same time (16:47~UT) as the time derivative of the SXR flux. 
The temporal correlation between HXR and SXR time derivative has been known as 
the Neupert effect \citep{Neupert68,Dennis93}, with the basic idea that non-thermal electrons
precipitate at the chromosphere, losing their energy by Coulomb collision to give rise to
HXR emission, and at the same time, driving chromospheric evaporation to fill
post-flare loops that are subsequently observed in SXRs.
This flare, however, exhibits a similar Neupert effect between the SXR derivative and UV, 
instead of HXR, light curves. This is evidence of continuous energy deposition in the lower 
atmosphere, most likely by thermal conduction, after the impulsive
phase when thick-target HXR emission is no longer significant.

The time sequence of UV images further shows that the continuous rise of the UV emission after the HXRs 
is produced by newly brightened UV ribbons (bottom panel in Figure~\ref{Fig1}). The spread of the flare UV
ribbons across the longitudinal magnetic field (lower right panel)
provides a measurement of magnetic reconnection flux \citep{Qiu02, Longcope07}, plotted in the top panel.
The reconnection flux starts to grow at 16:30~UT at the beginning of both UV and X-ray light curves.
The continuous increase of reconnection flux and dramatic decay of non-thermal emission after 16:50~UT 
(the post-impulsive phase) confirm that reconnection continues to form new loops and release energy 
in them, and the thermal process is dominant in this late phase of the flare. The post-impulsive 
reconnection flux amounts to 2$\times$10$^{21}$~Mx, about one third of the total reconnection flux measured for this flare.

Comparison of the UV and X-ray observations of the flare suggests that imaging UV observations
provide information of energy release, which is then transported to the lower atmosphere by either 
non-thermal electrons or thermal conduction, in newly formed flare loops anchored at newly brightened UV ribbons. 
Based on this idea, we have analyzed spatially resolved UV light curves, using them to construct
heating rates of flare loops in a C3.2 flare \citep{Qiu12}. The analysis is applied to this flare, as well,
in the following Sections.

\subsection{Characteristics of UV Light Curves}\label{Sec2.2}
\begin{figure}
\epsscale{0.95}
\plotone{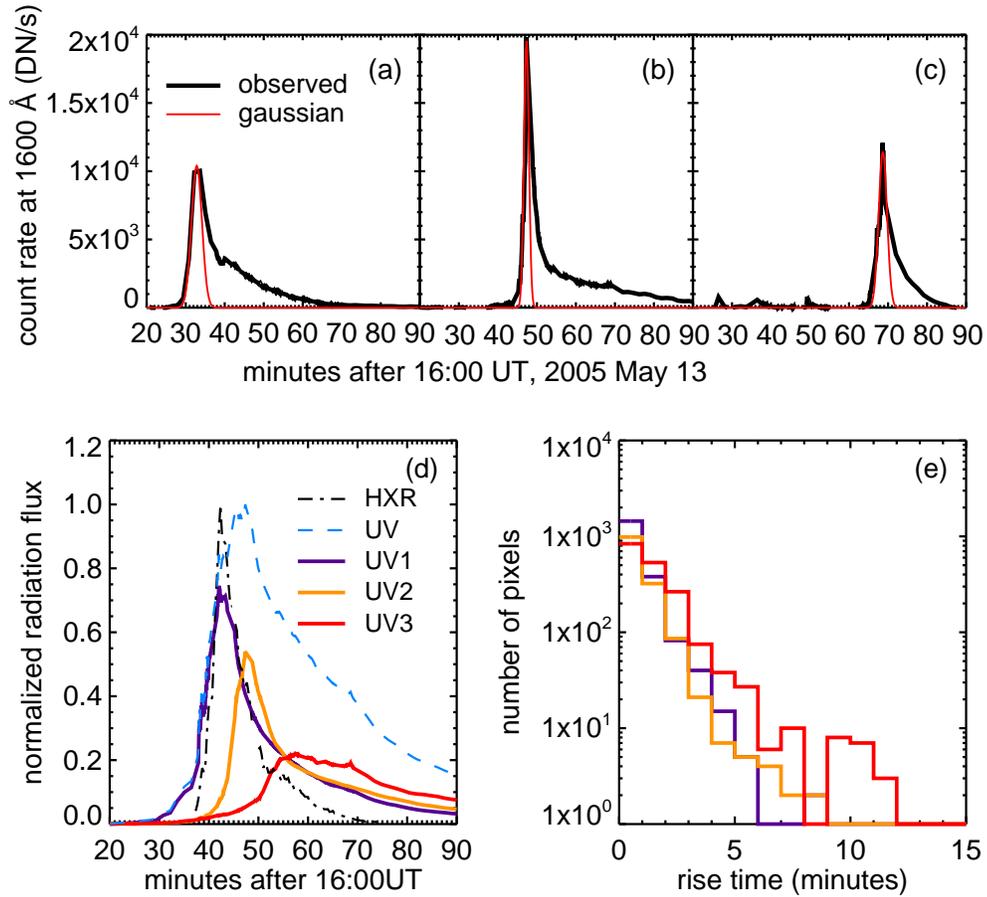}
\caption{Upper: observed UV 1600~\AA{} light curves (thick black line) of three different flaring pixels, superimposed with
the Gaussian function (thin red line) which fits the impulsive rise of each individual light curve.
Bottom left: normalized background-subtracted UV light curve of all flaring pixels, and UV light curve of pixels 
peaking at three different stages (see text), compared with the HXR 25--50~keV light curve. The UV light curves 
are normalized to the maximum of the total UV light curve. Bottom right: histograms of the rise times of UV light curves 
for pixels brightened at three different stages as shown in the bottom left panel.
} \label{Fig2}
\end{figure}

{Examples of} the UV light curve in a flaring pixel (1\arcsec\ by 1\arcsec), shown in the top panel of Figure~\ref{Fig2},
typically exhibits a rapid rise, peaks within several minutes, and then decays slowly with a 
characteristic \textquotedblleft cooling\textquotedblright \ time of more than 10~minutes. Such characteristics
have been found in UV observations of other flares as well \citep{Qiu10, Cheng12, Qiu12}. Since 
the lower atmosphere (the transition region and chromosphere) responds to energy deposition on very short timescales,
the rise time of UV emission reflects the timescale of energy release in the flare loop anchored at this pixel.
The gradual decay, on the other hand, is coupled with subsequent cooling processes in the overlying corona. 

To characterize the rapid rise of the spatially resolved UV light curves, we
fit the rise of the UV count rate light curve (in units of DN s$^{-1}$) to a half
Gaussian,
 \begin{equation}I(t) = I_0 \exp\left[\frac{-(t - t_0)^2}{2\tau^2}\right],  (t \leqslant t_0) \label{eq_halfgauss} \end{equation}
where $I_0$ is the background-subtracted peak count rate, $t_0$ is the peak time, 
and $\tau$ is the characteristic rise time. For this flare, 5127 
flaring pixels (each of size 1\arcsec\ by 1\arcsec) are identified
from UV images with 3~s cadence. 
The lightcurve
of each pixel is smoothed to 10~s and its rise phase is fitted to a half Gaussian. Examples of the fit to 
the observed light curve are plotted on the top panel of Figure~\ref{Fig2}. 
Histograms of the rise times from fits to all the flaring pixels are given in the bottom 
panel of Figure~\ref{Fig2}, showing that most flaring pixels rise over timescales from
a few tens of seconds to a few minutes. These rise times are systematically shorter than
those derived by \citet{Qiu12} in another C3.2 flare. This
may be due to the higher cadence (3~s
) of the observations in this flare compared with the 30~s cadence of the other data, 
or more likely, the more impulsive nature of this flare that has significant non-thermal emission.

Observations also show that UV pixels peaking after 16:52~UT, when the HXR emission has been reduced significantly,
tend to evolve more slowly than those peaking earlier. This suggests different evolutionary 
timescales in both the heating and cooling
phases when energy is transported through flare loops by different mechanisms, by non-thermal electrons or by thermal 
conduction. To examine the distinction, we compare UV light curves of pixels peaking 
in three stages relative to the HXR evolution. Pixels in the first stage (UV1) peak before 16:45~UT when HXR
emission is significant. These pixels are marked in dark black to light blue colors in Figure~\ref{Fig1}, and the sum
of their light curves ( purple 
curve in Figure~\ref{Fig2}d) has a very good temporal correlation with the HXR light curve.
UV pixels categorized in the second stage (UV2) peak during the HXR decay from 16:45~UT to 16:52~UT. The sum
of their light curves ( orange 
in Figure~\ref{Fig2}d) contributes mostly to the 
peak of the total UV light curve. UV pixels in the third stage (UV3; red 
in Figure~\ref{Fig2}d) peak after 16:52~UT, when $\ge$25~keV HXR emission has nearly ended. Compared with UV1 and UV2, 
the UV3 light curves evidently rise more slowly and also decay more gradually. In Figure~\ref{Fig2}e, 
we plot the histograms of the rise times for pixels at the three stages, which shows that UV3 
pixels rise more slowly on average. The differences of the UV light curves in the three stages 
indicates the different effects of beam heating and conductive heating on UV emissions, and in turn, the
UV evolution could give us a clue to the nature of energy release, thermal or non-thermal. 

Hydrodynamic and radiative transfer models of the lower atmosphere during flares
have shown that the lower atmosphere responds within a few seconds to the onset of impulsive
energy injection \citep{Emslie85, Fisher85_6, Canfield87}. Therefore, the rapid rise of UV emission from the upper atmosphere
or transition region may be considered to scale with the impulsive energy release in reconnection-formed
flare loops. With this idea, we have implemented a method to construct heating functions and used them 
to compute plasma evolution in flare loops for a C3.2 flare, which is primarily a thermal flare \citep{Qiu12}. 
In this study, we apply the same method, using the rising UV emission at the foot-point as an 
indicator of energy injection into the newly formed flare loop (or flux tube) rooted at the foot point. 
We take the start-time of the UV brightening as the onset of the reconnection event forming the new flux tube. 
The rise time of the UV brightness gives the duration of the impulsive energy release in the newly 
formed tube, and the maximum brightness of the pixel reflects the magnitude
of the energy release (or heating rate) in the flux tube; this simply assumes that a brighter pixel
is more strongly heated. These observationally measured quantities may then be used to construct
heating rates and study subsequent plasma evolution inside flare loops that are formed and heated
sequentially during the flare. In constructing the heating functions, significant progress in the 
present study is inclusion of the heating term by non-thermal particles that are evident during
the impulsive phase.

\section{Modeling Plasma Evolution in Flare Loops}\label{Sec3}

We have identified over 5000 brightened pixels of size 1\arcsec\ by 1\arcsec\ in {\it TRACE} 1600~\AA{} images, and assume
that a half loop (or flux tube) of constant cross-section (1\arcsec\ by 1\arcsec) is anchored at each of the pixels. 
These are half loops, because we do not identify connectivity between positive and negative foot-points. 
For each half loop, we compute the time-evolution of plasma density, temperature, and pressure averaged along the loop using 
the EBTEL model.
The energy input term in the model, the heating rate, is constructed from observed UV count rates at the foot point.
The time-dependent differential emission measure (DEM) is then  derived from these 5000 half loops, 
which will be used to compute SXR flux and compare with observations by {\it RHESSI} and {\it GOES} in
Section~\ref{Sec4}.

\subsection{Loop Evolution via EBTEL}\label{Sec3.1}

The 0D EBTEL model calculates mean properties of loop plasmas, 
which have been shown to reasonably agree 
with mean values from simulations using the 1D hydrodynamic code called Adaptively Refined Godunov Solver \citep[ARGOS;]
[]{Antiochos99}. The 0D model is highly efficient at computing 
plasma evolution for over 5000 half loops in our study 
of the M8.0 flare. The EBTEL model solves two equations. The energy (or pressure) equation takes into account the prescribed
heating rates ($H$) as the energy input term and coronal radiative loss ($R_c$) as well as the total loss ($R_{tr}$) 
through the base of the loop (so called transition region) as energy loss terms. The mass equation is governed
by mass flow between the transition region and the corona. In the EBTEL model, this flow is a result of the difference 
between the energy input, including the beam heating and conductive flux ($F_0$) from the corona, 
and the total loss ($R_{tr}$) at the base (transition region). During the heating phase, the energy input to the 
transition region dominates the loss term, driving upflows known as chromospheric evaporation. During the decay, 
the coronal plasma in the loop is cooling through thermal conduction and radiation; meanwhile, the loss 
through the transition region exceeds the conductive flux into it, which drives downflows, called coronal condensation. 

The mean coronal electron density ($n$) and pressure ($P$) in each loop 
evolve according to the EBTEL equations,   
\begin{eqnarray}
  \frac{dn}{dt} &=& -\frac{c_2}{5c_3k_{\rm B}T}\,\left( \frac{F_0}{L} + c_1n^2\Lambda(T)
  -\frac{\Gamma(t)}{L} \right) \label{EBTEL_eq1} \\ 
  \frac{dP}{dt} &=& \frac{2}{3}\,\left[ Q(t) - (1+c_1)n^2\Lambda(T) 
  + \frac{\Gamma(t)}{L} \right]\label{EBTEL_eq2}
\end{eqnarray}
The mean temperature $T$ is determined by the 
ideal gas law, $P = 2nk_{\rm B}T$ (including both electrons and ions).  
In the equations, $k_{\rm B}$ is Boltzmann's constant, $c_1$ is the ratio of the loss 
through the transition region to the coronal radiation ($R_c/R_{tr}$), $c_2$ is the ratio of the average 
coronal temperature to the apex temperature, and $c_3$ is the ratio of the coronal base temperature to the apex temperature.
With the symmetry assumption, EBTEL only models heating of a half loop, with
$L$ being the length of the half-loop. We prescribe two heating terms in the equation: 
the {\it ad hoc} volumetric heating rate, $Q(t)$, and  
the energy flux carried by beam-driven upflows, $\Gamma(t)$. The coronal radiative 
loss is given by $R_c = n^2 \Lambda(T)$, $\Lambda(T)$ being the empirically-determined
radiative loss function for optically-thin plasmas \citep[see equation (3) of][for details]{Klimchuk08}.
$F_0$ is the conductive flux at the base of the corona, which is defined by \citet{Klimchuk08} as 
the location where thermal conduction changes from being an energy loss above to an energy source below. 
The classical form of conductive flux is used \citep{Spitzer1962}, 
\begin{equation} F_0 = -\kappa_0 T^{5/2} \frac{\partial T}{\partial s} \approx -\frac{2}{7} \kappa_0\frac{(T/c_2)^{7/2}}{L} \label{eq_spitzer} \end{equation}
where $\kappa_0$ is the thermal conductivity coefficient, 
taken to be $1.0 \times 10^{-6}$ in cgs units. In the EBTEL model, $F_0$ is saturated for large temperature gradients; 
in the latest version of EBTEL \citep{Cargill2012_2, Cargill2012_3}, the gravity is included 
in calculating of $c_1$ for semi-circular loops, while the dependence of $c_2$ and $c_3$ on gravity is negligible.

There are three parameters in the EBTEL models, $c_1$, $c_2$, and $c_3$.
In the latest version of EBTEL \citep{Cargill2012_2, Cargill2012_3}, which is used in the
this study, $c_1$ is self-consistently determined by plasma properties inside the loop. 
Its value varies about the mean value 2.1 during flux tube evolution, which is not very sensitive to
different heating rates in different flux tubes. Fixed values of $c_2 = 0.87$ and $c_3 = 0.5$ 
are used in the study. These are mean values determined from 1D simulations. In the simulations, $c_2$ and $c_3$ 
usually change during the loop evolution, but only within a small range. For simplicity, 
in this study, we use fixed mean values of $c_2$ and $c_3$ for 5000 loops throughout their evolution, 
considering that differences produced by using varying values of these parameters will become insignificant 
when we sum up contributions from 5000 loops.

The critical input to the EBTEL model, also a focus of our present study, is the heating rate. In general, 
it includes two parts. The first contribution, denoted by an {\it ad hoc} volumetric heating rate $Q$ (in units of erg~cm$^{-3}$~s$^{-1}$), 
is by {\it in situ} heating in the corona. It may result from current dissipation, shocks \citep[e.g.,\ ][]{Longcope10,Longcope11}, 
electrons trapped and scattering in the corona\citep[e.g.,\ ][]{Somov97,Karlick04,Caspi10}, or even return current 
\citep[e.g.,\ ][]{Knight77, Emslie80, Holman12}. In the present study, the exact mechanisms 
for this terms are not discussed. The second contribution, denoted as $\Gamma$ (in units of erg~cm$^{-2}$~s$^{-1}$), 
is coronal heating {\em from the lower atmosphere} due to evaporation driven by non-thermal electrons
that precipitate at the lower atmosphere during the impulsive phase. Of the total flux carried by a non-thermal beam, 
$\Gamma$ is a fraction of that sent back upward into the corona.
These two terms are distinguished as they heat the corona in different ways, $Q$ by primarily raising the temperature of 
the coronal plasma, and $\Gamma$ by primarily raising the density - therefore the $\Gamma$ term also enters the density equation. 
Predictably, the different heating styles result in different evolutionary 
patterns. Furthermore, these two terms play different roles in different stages of the flare. Specifically, 
the $\Gamma$ term as driven by beams is included only during the impulsive phase when thick-target HXR emission is evident.

We must note that thermal conduction alone is able to produce chromospheric evaporation, which, in the EBTEL model,
is the consequence of coronal evolution when the conductive flux exceeds the loss through the lower
atmosphere during the heating phase, and is therefore not treated as an additional coronal heating term. 
The $\Gamma$ term, on the other hand, is considered in this study to be produced by particles that 
instantaneously heat the lower atmosphere, and drive upflow {\em independent of} the coronal situation.
Therefore, this term contributes to both energy and density of coronal plasmas.

For the 2005 May 13 flare, {\it RHESSI} observations show that there is significant HXR emission
with energy up to 300~keV during 
the first ten minutes of the flare. So both {\it in situ} heating 
in the corona and heating by beam driven upflow are considered in this paper. Until now it has proven 
difficult in general to distinguish these two contributions, either theoretically or observationally.
In this study, we use an empirical method to introduce these two terms as scaled with the UV and HXR 
light curves. Determination of these two heating terms and the effect of the partition between 
the two will be discussed in the following text.

\subsection{Constructing Heating Functions in Flare Loops}\label{Sec3.2}
Following \citet{Qiu12}, we assume that the heating rate in a flare loop (or a flux tube) 
is proportional to the rise of the UV light curve at its foot-point, given as: 
\begin{equation}H_i(t) \equiv Q_i(t)L_i+\Gamma_i(t) = \lambda I_i\exp{\left[ -\frac{(t-t_i)^2}{2\tau_i^2} \right]} {\text{~erg~s}^{-1}\text{~cm}^{-2}};\ (0 < t < \infty)\label{eq_heating}\end{equation}
$H_i$ is the total heating flux in the loop anchored at $i$th pixel,  
which is composed of the {\it ad hoc} heating rate ($Q_iL_i$) and the beam-driven flux ($\Gamma_i$). 
Although we only fit the rise of the UV light curve to a half Gaussian, we consider the heating function to be  
symmetric, or a full Gaussian. Assuming semi-circular post-flare loops, we estimate the length of the $i$th half loop 
by $L_i=\frac{\pi}{2}D_i$, where $D_i$ is the distance of the foot-point to the polarity inversion line. 
As observed, flare ribbons expand away from the polarity inversion line in a rather organized manner; 
therefore, $L_i$ approximately grows linearly with the time of flare brightening ($t_i$). For the 5000
half loops, $L_i$ ranges from 35 to 55~Mm. Although loops are in general not semi-circular, our experiments
have shown that variations of the loop lengths within a factor of 1.5 do not significantly change
the synthetic total emission.  

To relate the total heating rate to UV emission, we employ a scalar 
$\lambda$ that converts 
the count rate (DN s$^{-1}$) to a heating rate (erg s$^{-1}$). The value of this parameter 
depends critically on the lower atmospheric response to beam heating or conductive heating 
and on mechanisms of UV emission. As a rule of thumb, 
we consider that $\lambda$ takes a larger value when conduction heating dominates 
than when beam heating dominates, for the simple reason that, with the same amount of energy, 
beam heating occurring in
the lower atmosphere would generate stronger UV emission than conductive flux \citep[e.g.,][]{Emslie85}.
In this study, $\lambda$ takes the value of 1.9
$\times 10^5$~ergs~DN$^{-1}$ for loops 
whose foot-point UV light curves peak before 16:48~UT, then linearly increases until 16:52~UT; afterwards, 
when HXR emission has finished, $\lambda$ stays constant at 2.5
$\times 10^5$~ergs~DN$^{-1}$. 
Note that in this study, we do not model the lower atmospheric 
heating and dynamics, but instead use this simple empirical 
model to minimize the number of free parameters. The $\lambda$ values quoted here are  determined by best matching the  model-computed 
time-dependent SXR emissions with those observed by {\it GOES}, 
as will be described in the next section. These $\lambda$ values, combined with the peak count rates, 
correspond to the peak heating flux ($H$) of order $10^8$ to 10$^{10}$~ergs~s$^{-1}$~cm$^{-2}$ for the
few thousand half loops in our model.

In the heating term, the coronal {\it in situ} heating $QL$ is present in all flare loops; $\Gamma$ is present 
only for loops brightened during the impulsive phase (UV1 and UV2 pixels), and is gradually switched off 
after 16:52~UT (UV3 pixels). For one loop, the {\it ad hoc} heating rate and beam driven energy flux both have 
the same Gaussian time profile of $\exp{[-(t-t_i)^2/(2\tau_i^2) ]}$, and the partition of the 
beam driven energy flux, i.e.,\ $\Gamma_i/H_i$, is constant. For different loops heated at different times
represented by $t_i$, the partition $\Gamma_i/H_i$ is different; whereas the net heating flux $H_i$ in a loop is proportional to 
the UV count rate at the foot-point, the beam driven flux $\Gamma_i$ is assumed to be proportional to the 
$\ge$25~keV HXR light curve. This rather simplified treatment can be justified by the hydrodynamic 
simulations of chromospheric evaporation showing that the evaporation upflow roughly increases as the heating 
energy flux increases \citep{Fisher84, Fisher85_5} in the range of 10$^{9}$-10$^{10}$~erg~s$^{-1}$~cm$^{-2}$. 
The estimated heating flux of the majority of flux tubes in the impulsive phase in our study is 
within this range. During the impulsive phase, the HXR spectral index does not vary significantly 
(not shown in this paper), so we can approximate the total beam energy flux, and subsequently the energy flux in the beam
driven upflow, as proportional to the HXR count rate light curve.
The partition $\Gamma_i/H_i$, therefore, is time-dependent, and is empirically given by
$\Gamma_i/H_i = \gamma_m \eta(t_i)$,
where $\eta (t_i)$ is a time-dependent function that tracks the HXR light curve,
$\gamma_m = 0.4$ is the maximum partition of the non-thermal energy flux used in this study, and this
maximum partition occurs when HXR emission peaks at about 16:42~UT.  

In addition to the impulsive flare heating, a constant background heating rate of 
order $1\times10^{-4}\text{~ergs~s}^{-1}\text{~cm}^{-3}$, which is a few thousandths of the 
maximum heating rate constructed from UV observations, is imposed on each loop to produce an initial 
equilibrium of an average coronal temperature of $1.8$ MK and density of $6\times 10^8$~cm$^{-3}$.
\citet{Qiu12} have shown that neither the initial state of the loop nor the background heating will affect
the plasma evolution as soon as impulsive heating occurs, since the flare heating rate is a few orders 
of magnitude larger than the background heating rate.  

The so devised heating function in a single flux tube is mostly constrained by foot-point UV and HXR observations,
but also depends on two free parameters $\lambda$ and $\gamma_m$. $\lambda$ determines the total amount of
energy used to heat the corona, and $\gamma_m$ determines the non-thermal partition. Since heating 
via upflows or via direct heating leads to different plasma evolution
pattern, the resultant coronal radiation spectrum will differ. Our recent experiments with another flare 
have shown that model-predicted coronal radiation by high temperature plasmas is very
sensitive to parameters defining the heating function but rather insensitive to other model parameters
like $c_1$ \citep{Qiu12}. Therefore, these parameters will be eventually constrained
by comparing model-predicted and observed X-ray emission. In this study, we start the model with an 
initial guess of the parameters, and gradually adjust them so that the synthetic SXR spectrum and 
light curves best match those observed by {\it RHESSI} and {\it GOES}.

\subsection{Evolution of Flare Plasma in One Loop}\label{Sec3.3}
\begin{figure}
\epsscale{0.9}
\plotone{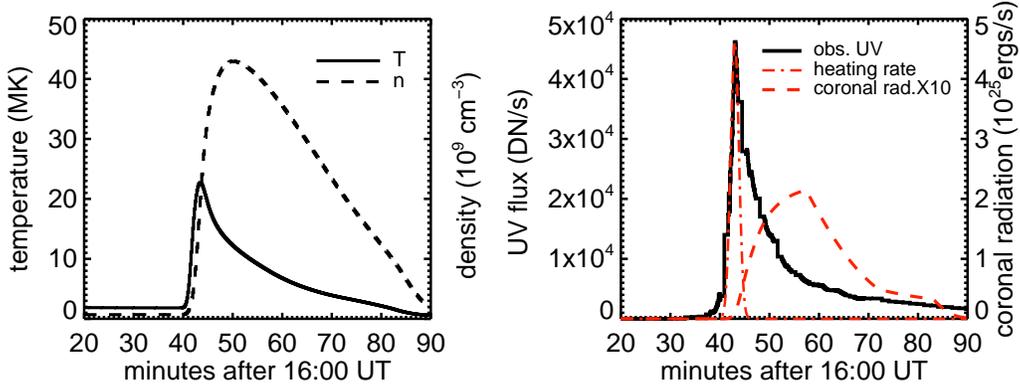}
\caption{Evolution of plasma properties in a flare loop with cross section 1\arcsec\ by 1\arcsec\ rooted
at one flaring pixel. Left: computed time profiles of coronal-averaged temperature (solid) and density (dashed) of the flare loop. 
Right: observed UV 1600~\AA{} count rate light curve (solid black line), the constructed heating function (dot-dashed red line), and 
computed coronal radiation rate (dashed red line) of the loop.} \label{Fig3}
\end{figure}

With methods described above, we compute plasma evolution in each flux tube anchored at the UV flaring pixel
using the EBTEL model. Figure~\ref{Fig3} shows, in the left panel, time profiles of the mean temperature (solid) and density (dashed) 
of a single flux tube rooted at its UV foot point with constant cross-sectional area 1\arcsec\ by 1\arcsec.
It is heated impulsively by a Gaussian-profile heating rate (red dot-dashed line in the right panel) constructed 
from the UV light curve (black solid line in the right panel) with $\tau = 50$~s, and $H_{max} = 9 \times 10^{9}
\text{~ergs~s}^{-1} \text{~cm}^{-2}$, of which $40$\% is carried by the beam-driven upflow. 
The $\lambda$ and $\gamma_m$ values used here are determined from comparing synthetic SXR light curves 
and spectra with observations, which will be described in the following sections.

Coronal temperature rises as the flux tube in the corona is heated by {\it ad hoc} heating. At the same time, 
the lower atmosphere is heated by non-thermal electron beams and conductive flux, giving rise to enhanced UV emission
as observed by {\it TRACE}. Chromospheric evaporation is driven to fill the coronal loop. An 
increase in coronal density leads to enhancement of coronal radiation (red dashed line in the right panel), 
which then cools the plasma in the flux tube. The coronal temperature begins to decline immediately after the peak of 
the heating, but the density continues to grow until the energy loss in the transition region exceeds the conduction 
flux, causing a downward flux, or coronal condensation. 
The coronal density and radiation then decline, and the flux tube experiences a long decay. Notably, 
the observed UV 1600~\AA\ emission decays on the same timescale of coronal evolution. This has been previously known as
optically-thin transition region lines, such as the resonant C~{\sc iv} line that
dominates the 1600~\AA\ broadband UV emission, behave as a ``coronal pressure gauge" \citep{Fisher87, Hawley92}. 

\subsection{Effect of Beam Driven Flux on Plasma Evolution}\label{Sec3.4}
\begin{figure}
\epsscale{.9}
\plotone{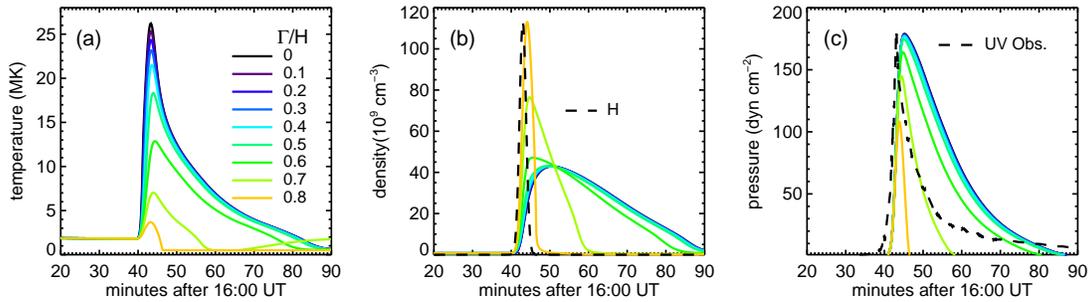}
\caption{Evolution of coronal-averaged temperature (a), density (b), and pressure (c) of a flux tube 
with varying fraction of energy flux carried by beam driven upflows, $\Gamma/H = 0 \textrm{--} 0.8$. Also plotted
in dashed lines are arbitrarily scaled total heating rate in panel (b) and observed UV light curve in panel (c).
} \label{Fig4}
\end{figure}

To determine $\gamma_m$, we investigate the effect of beam driven flux in the EBTEL model. 
We examine the evolution of coronal plasma in the same flux tube shown in Figure~\ref{Fig3} with the same impulsive heating rate 
$H_{max} = {9 \times 10^{9} 
\text{~ergs~s}^{-1} \text{~cm}^{-2}}$ but with varying partition, i.e.,\ $\Gamma/H$ values.
The time profiles of coronal-averaged temperature, density, and pressure with different $\Gamma/H$ 
values are plotted in Figure~\ref{Fig4}. 

It is shown that, in general, a greater amount of beam driven flux results in lower temperature and higher density.
If the beam driven flux does not dominate, i.e.,\ $\Gamma/H < 0.5$, the average temperature decays
to $10$ MK in a few minutes, and then continues to decay toward the pre-flare value. From about 
ten minutes after the heating, the decay of coronal temperature and density is nearly identical for different $\Gamma/H$ values. 
The same evolution pattern of the apex density and temperature is also displayed in 1D hydrodynamic simulations 
by \citet{Winebarger04}, who modeled loop heating with thermal conduction. Their study showed that a flare
loop impulsively heated by the same amount of the total energy, but with varying magnitude and duration and at different
locations along the loop, would reach the same equilibrium point when radiation and conduction are comparable,
and the evolution of the apex density and temperature is identical thereafter.

On the other hand, when the beam driven flux dominates the energy budget with $\Gamma/H > 0.5$, the flux tube
will attain much higher density in a short time due to strong upflow. Meanwhile, less direct heating
$Q$ results in a lower peak temperature. The two effects, high density and low temperature, 
lead to faster
radiative cooling in the decay phase. Therefore, the flux tube evolves on much shorter timescales. A similar evolution pattern has also been produced in the 1D numerical simulation that takes into account
beam heating with different pitch angles \citep{Reeves12}.

\begin{figure}
\epsscale{.7}
\plotone{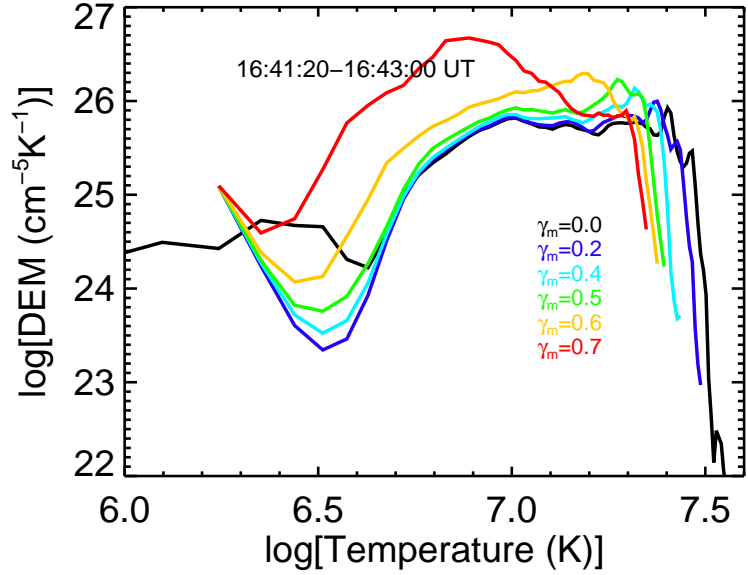}
\caption{The differential emission measure (DEM) derived from multiple flare loops from 16:41 to 16:43~UT, 
with different $\gamma_m$ but the same total heating rates.
} \label{Fig_dem_nth}
\end{figure}

The experiment above suggests that varying the partition parameter $\Gamma/H$ will lead to different temperature and density
in the heating phase, and for very large $\Gamma/H$ values, evolution of the flux tube in the decay phase is also
modified significantly. Therefore, the experiment provides us with two ways to estimate the optimal $\Gamma/H$ value
in this flare. First, we may compare the synthetic SXR spectrum using the DEM from modeling over 5000 flux tubes
with the X-ray spectrum observed by {\it RHESSI} during the heating phase. Figure ~\ref{Fig_dem_nth} shows the DEM constructed using
different $\gamma_m$ values for the 5000 tubes. It is seen that, given the same amount of total energy, 
the larger value of $\gamma_m$, i.e., the larger amount of energy carried by beam driven upflow, will 
lead to greater DEM increase at lower temperatures, whereas the DEM at the high temperature end would be reduced.
Such variation will affect the synthetic SXR spectrum formed at the temperature range of a few to a few tens
of MK. Therefore, the value of $\gamma_m$ can be estimated by comparing the synthetic SXR spectra with {\it RHESSI} 
observations, as will be presented in Section~\ref{Sec4.2}.

Second, we will compare the observed and modeled timescales of the foot-point UV emission during the decay phase 
of the flare, based on the principle that UV line emission behaves as a coronal pressure gauge in the decay phase 
\citep{Fisher85_6, Fisher87, Hawley92}. These will be elaborated on in Section~\ref{Sec4.3}. As a quick look, the 
right panel of Figure~\ref{Fig4} shows the observed UV light curve versus pressure; it appears that the UV 
emission decays on the same timescale as the coronal pressure for $\Gamma/H < 0.5$. 

\subsection{Properties of Coronal Plasmas in Multiple Flare Loops}\label{Sec3.5}
\begin{figure}[h]
 \begin{center}$
 \begin{array}{cc}
 \includegraphics[width=2.in]{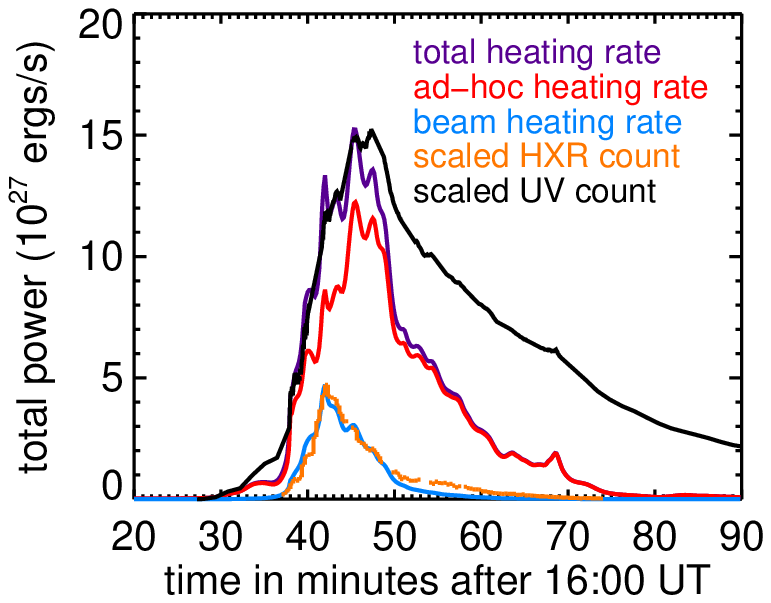} &
 \includegraphics[width=4.in]{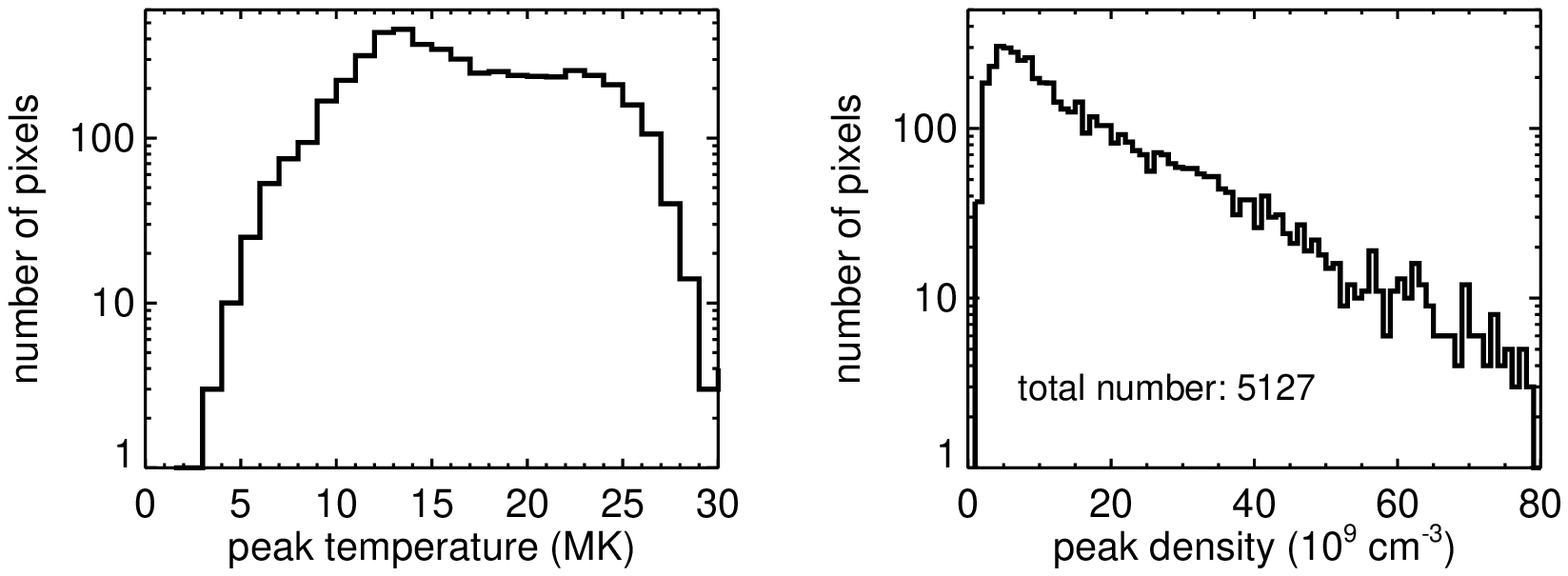} 
 \end{array}$
 \end{center}
 \caption{Left: the sum of the total heating rate (purple), 
ad hoc heating rate (red), 
and beam heating rate (blue) 
in all flare loops, in comparison with the observed total UV light curve (black
) and HXR 25--50~keV light curve (orange), 
both arbitrarily scaled. Right: peak temperature and density 
distributions in over 5000 modeled flare loops with lengths in the range of 35--55~Mm and cross-sectional area
of 1\arcsec\ by 1\arcsec.}\label{Fig6}
\end{figure}

Our methods yield the best estimate of $\lambda$ and $\gamma_m$, with values given in Section~\ref{Sec3.2}, 
with which we can determine heating functions of over 5000 flux tubes. 
The left panel of Figure~\ref{Fig6} shows the time profiles of the total {\it in situ} heating rate ($ \sum Q_iL_i$) and total
beam heating rate ($\sum \Gamma_i$) summed for all the flux tubes, in comparison with the 
observed UV total counts light curve and HXR 25-50~keV counts light curve. The total heating rate ($\sum H_i$) is shown nearly 
scaled with the total UV count rate light curve from the rise to the peak -- note that
the decay of the UV light curve is governed by coronal evolution, which is not part of the heating. The
beam driven energy flux is proportional to the observed $\ge$25~keV HXR light curve. The time integral
of the total heating rate and beam heating rate yields the estimate of the total energy used to 
heat the corona and the total energy carried by beam-driven upflows in this flare, which are $1.22
\times10^{31}$ ergs and $1.9
\times10^{30}$~ergs, respectively. The figure
shows that the largest non-thermal partition amounts to about 40\%, which occurs around the peak of
HXR emission at 16:42~UT. Note that although the total heating flux $H_i$ in a single flux tube is considered 
to be proportional to the UV light curve at the foot-point, the sum of the total heating fluxes in all flux tubes,
$\sum H_i = \sum (Q_iL_i + \Gamma_i$), is scaled with the total UV light curve by different values of $\lambda$ during
different phases of the flare, as can be seen from the Figure. $\lambda$ is time-dependent in our method; given 
the same amount of energy, beam heating produces greater UV emission compared with conduction heating \citep{Emslie85}. 

With these heating rates, we compute plasma evolution in over 5000 flare loops formed by reconnection and heated 
at different times and by different amounts of energy. The distribution of the peak 
temperatures and densities of these flux tubes is shown in the right panels of Figure~\ref{Fig6}. For this flare, 
the peak temperature of most flux tubes varies from 7 to 26~MK, and the peak temperature distribution
appears bi-modal. The second peak in the distribution at around 22~ MK is mostly contributed 
by flare loops formed around 16:45~UT, and is close to the effective temperature derived from the ratio of the 
two-channel (1--8~\AA{} and 0.5--4~\AA{}) emission measured by {\it GOES} with the isothermal assumption \citep{White05}.
The peak density of the flux tubes ranges from  5--30$\times 10^9$~cm$^{-3}$. 

\begin{figure}
\epsscale{.7}
\plotone{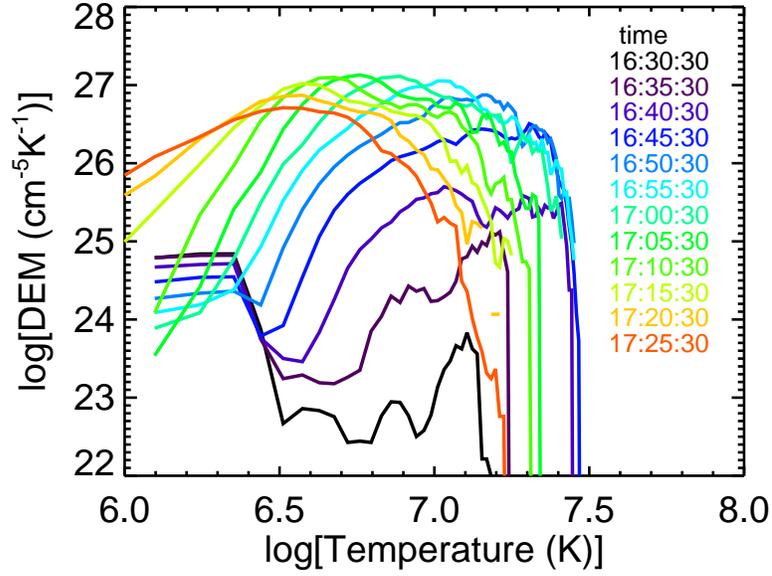}
\caption{Time evolution of the coronal DEM from the model of the 2005 May 13 flare. The high emission measure at
low temperature at the beginning of the flare is produced by background heating.} \label{Fig7}
\end{figure}

The DEM of the coronal plasma  is readily calculated from the 
temperatures and densities of these loops. Figure~\ref{Fig7} shows the time evolution of the coronal plasma DEM 
averaged every 20 seconds for the M8.0 flare. Note that the higher DEM at about 2 MK at the beginning 
of the flare is due to background heating. It is seen that in the rise phase of the flare, the DEM increases 
toward higher temperature, suggesting that more flux tubes are heated to higher temperatures, while during the 
decay phase, the DEM rises at lower temperature, reflecting cooling of the flux tubes. The time evolution of the coronal 
DEM will be used to calculate time-dependent SXR emission to verify the model results,  as discussed in the next Section.

\section{Comparison with Observations}\label{Sec4}

The time-dependent coronal DEM computed from over 5000 flux tubes is convolved with 
instrument response functions to compute synthetic SXR light curves 
and spectra, which are compared with those observed by {\it GOES} and {\it RHESSI} (Sections~\ref{Sec4.1} and ~\ref{Sec4.2}, respectively). 
The transition region DEM during the decay phase is also derived from the coronal pressure \citep{Fisher85_6, Fisher87, Hawley92},
with which we calculate the the C~{\sc iv} line emission using the CHIANTI atomic database 
\citep{Dere97, Landi12}, and compare it with UV 1600~\AA{} observations (Section~\ref{Sec4.3}). 
 
\subsection{Comparison with {\it GOES} Light Curves}\label{Sec4.1}
The two X-ray Sensor (XRS) photometers onboard {\it GOES} measure full-disk integrated SXR emissions in two energy bands, 
1--8~\AA{} and 0.5--4~\AA{}, with 3-second cadence. The solar 1--8~\AA{} flux, which is dominated by continuum emission 
\citep{Mewe72, Kato76}, is believed to originate from hot ($T>10^6$~K) plasma in coronal loops in active regions. 
The full-disk SXR images observed by the Solar X-ray Imager \citep[SXI;][]{Hill05, Pizzo05} 
onboard {\it GOES} indicate that NOAA Active Region 10759, where the M8.0 flare occurred, is the dominant SXR source 
on 2005 May 13 from 16:00~UT to 20:00~UT (images not shown here). Therefore, the background-subtracted {\it GOES} light curves 
reflect SXR emissions by the M8.0 flare. The comparison of the synthetic SXR light curves 
from the modeled flare plasma with the observations by {\it GOES} XRS yields the best-fit model parameter $\lambda$.

\begin{figure}
\epsscale{.7}
\plotone{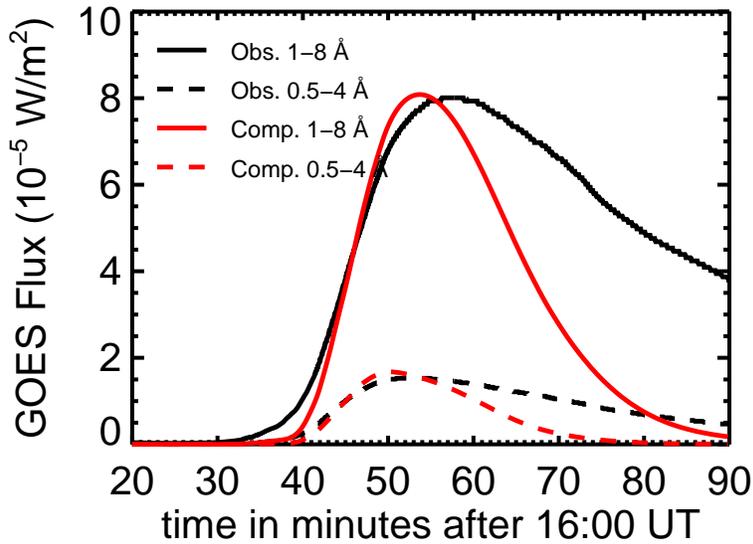}
\caption{Left: 
Comparison of the computed (red) 
and {\it GOES} observed (black) 
SXR fluxes at 0.5--4~\AA{} (dotted) and 1--8~\AA{} (solid). 
} \label{Fig8}
\end{figure}

Figure ~\ref{Fig8} shows the synthetic SXR emissions in two channels plotted against the observed
light curves. In both band-passes, the synthetic light curves follow the observed ones
very well from the rise until 16:50 UT.
In the 0.5--4~\AA{} channel, the modeled and observed fluxes start to 
decline at around the same time, but the modeled flux decays more rapidly than the observed. 
In the 1--8~\AA{} channel, the modeled radiation flux begins to drop while the observed radiation 
continues to rise for another 5 
minutes. We also note that at the start of the flare before 16:40~UT, the
observed {\it GOES} 1--8~\AA{} flux starts to rise earlier than the modeled flux. As discussed
by \citet{Qiu12}, simply changing model parameters cannot compensate for the flux deficiency; instead, 
these discrepancies are most likely caused by weak heating events in the very early and late phases of the flare, which
might not be identified in the UV foot-point emission.

\subsection{Comparison with {\it RHESSI} Soft X-ray Spectra and Light Curves}\label{Sec4.2}
{\it RHESSI} is designed to observe solar high-energy emissions from SXRs to gamma-rays  (3~keV up to 17~MeV) with
an unprecedented combination of high time, spatial, and spectral resolutions \citep{Lin02}. 
Whereas the thick-target HXR ($\gtrsim$20~keV) observations provide us with a guide for constructing heating functions,
we will further compare the model computed SXR  ($\lesssim$20~keV) spectrum with that observed by {\it RHESSI}. The 3 to $\lesssim$20~keV X-ray emission
is usually thermal bremsstrahlung radiation produced by plasmas in the flare loops heated to  temperatures of
a few to a few tens of MK. {\it RHESSI} images confirm that this emission is generated in the corona.

We calculate the time-dependent SXR spectrum using the DEM distribution obtained from our 
model (see Figure~\ref{Fig7}). At each time (20~s cadence), we first calculate the X-ray spectrum observed at Earth
(in units of photons~cm$^{-2}$~s$^{-1}$~keV$^{-1}$) from the optically-thin thermal bremsstrahlung 
radiation using CHIANTI \citep{Dere97, Landi12}, including both the line and continuum contributions and assuming solar coronal abundances \citep{Feldman92} and ionization equilibrium \citep{Mazzotta98}.
The calculated spectrum is then convolved with the {\it RHESSI} detector response \citep{Smith02},
obtained from pre-flight and in-flight instrument calibration and modeling,
to convert the photon spectrum to the count spectrum that would be directly observed by {\it RHESSI}. The detector response
accounts for instrumental effects due to pulse pileup, escape of K-shell fluorescence photons from the detectors, attenuation by the aluminum shutters,
and the energy-dependent detector effective area, and includes the additional in-flight corrections from \citet{CaspiPhDT}.
To get the best comparison of the spectra, only detector G4 is used since it had the best in-flight 
resolution at the time of the flare \citep{Smith02}. By integrating the synthetic spectrum over different photon energy ranges, we also obtain synthetic SXR light curves in units of counts rate. 

\begin{figure}
\epsscale{.7}
\plotone{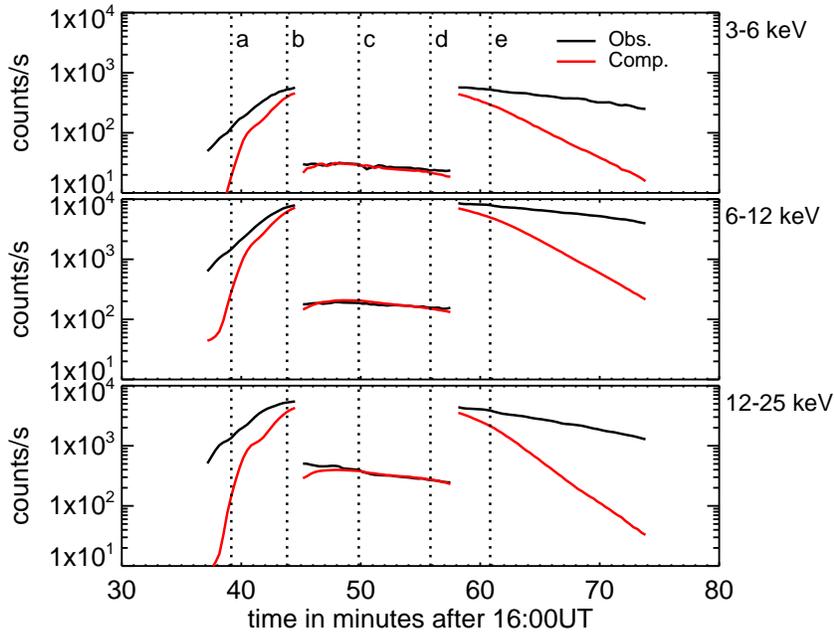}
\caption{Comparison of the synthetic SXR light curves from the model (red) 
with {\it RHESSI} observations (black) 
in the 3--6~keV, 6--12~keV, and 12--25~keV bands. The vertical dotted lines and letters a--e indicate the times when 
the spectral comparison is plotted in Figure~\ref{Fig10}. The discontinuities in the light
curves at 47~min and 57~min are due to changes of the {\it RHESSI} attenuator state, when the thick shutter
is inserted and later removed. }\label{Fig9}
\end{figure}

Figure~\ref{Fig9} shows the synthetic SXR light curves in the 3--6, 6--12, and 12--25~keV bands, respectively,
in comparison with {\it RHESSI} observations. For most of the impulsive phase, the synthetic
light curves agree very well with the observed ones. The synthetic fluxes are lower than observed
at the start of the flare, for a few minutes after 16:37~UT, when {\it RHESSI} emerges from eclipse. They then 
rise rapidly to catch up with the observed fluxes as more flare loops formed and heated by reconnection 
are identified in the UV foot-point observation. The computed and observed fluxes overlap
for 15 minutes from 16:44~UT until 16:59~UT, well after the impulsive phase. Afterwards,
the computed fluxes start to drop faster than observed. This 
is consistent with the model-observation 
comparison of the {\it GOES} light curves, which usually yield slightly lower plasma temperatures than RHESSI light curves \citep[e.g.,][]{Hannah08,Caspi10}. 
 
\begin{figure}
\epsscale{.7}
\plotone{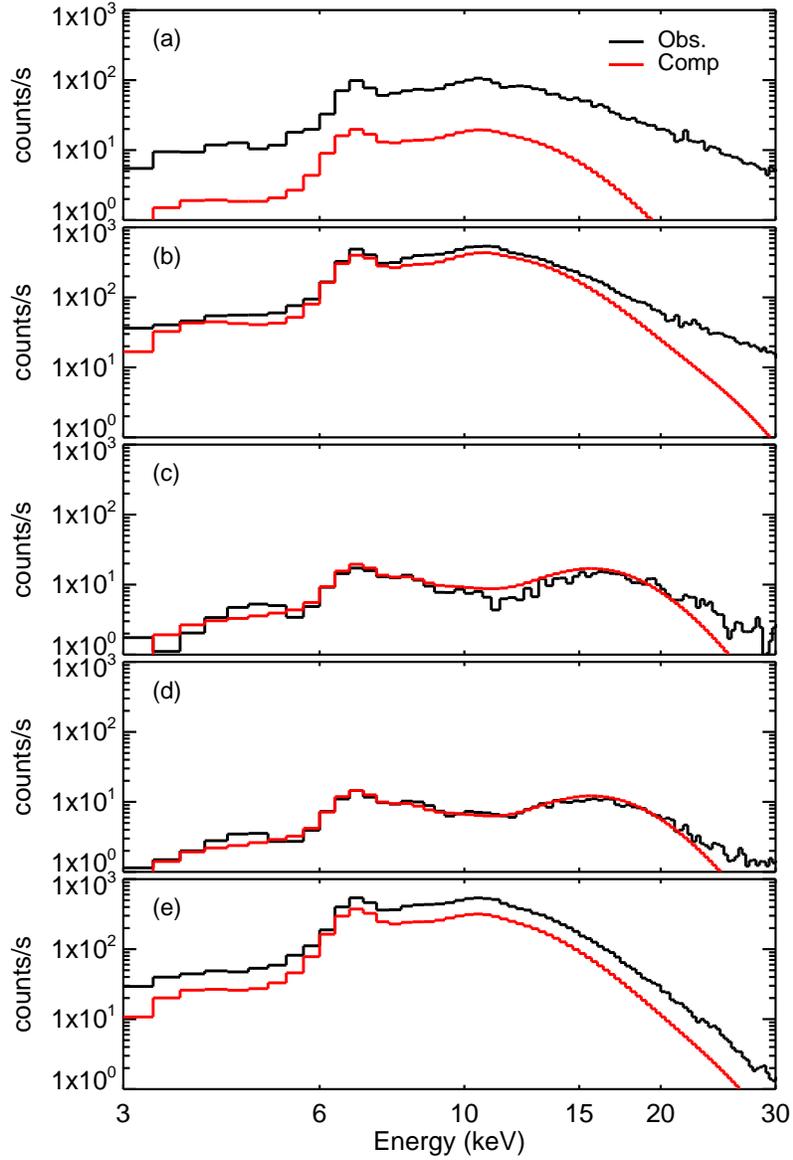}
\caption{ Comparison of synthetic SXR spectra from the model (red) 
with the {\it RHESSI} observation (black) 
at a few times during the flare. The time of each panel is indicated by vertical dotted lines in Figure~\ref{Fig9}.
For a complete view, a movie is available \href{http://solar.physics.montana.edu/wjliu/20050513/hsi_cmp_ebtel2.gif}{online}, 
which compares the observed (black) and computed (red)
spectra from 16:37 to 17:15~UT, with 20~s cadence. } \label{Fig10}
\end{figure}

\begin{figure}
\epsscale{.76}
\plotone{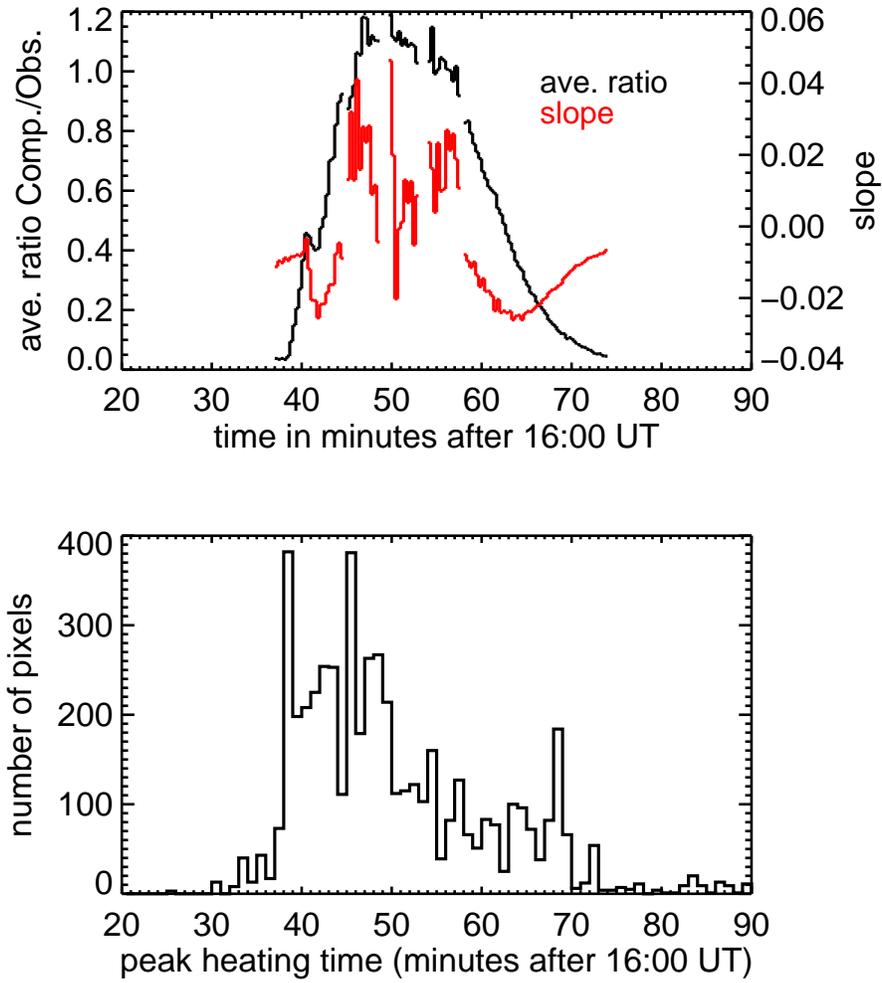}
\caption{Top: time profiles of the ratio (black) 
of the modeled SXR spectrum to that observed by {\it RHESSI} averaged over the 6--15~keV band,
and slope (red) 
of the ratio versus photon energy in the 6--15~keV band. Bottom: time histograms 
of the number of flaring pixels identified from the UV foot-point emission. } \label{Fig11}
\end{figure}

More details can be learned by comparing the SXR spectra. Figure~\ref{Fig10} illustrates the synthetic and 
observed spectra at a few different times, indicated in the top panel of Figure~\ref{Fig9}, during the flare.
(For a complete view, a movie comparing the modeled and observed SXR spectra, from 16:37 to 17:15~UT with 20~s cadence, 
is available \href{http://solar.physics.montana.edu/wjliu/20050513/hsi_cmp_ebtel2.gif}{online}.)
Figure~\ref{Fig11} summarizes the comparison of the time-dependent spectra with two quantities, the 
ratio of the synthetic to the observed counts, $\rho(\epsilon) = C_m(\epsilon)/C_o(\epsilon)$, averaged over the photon energy range 
$\epsilon$ from 6 to 15~keV, and the slope $\alpha$ of this energy dependent ratio versus the photon energy,  
which is obtained by fitting the ratio to a linear function of the photon energy $\rho(\epsilon) = \alpha \epsilon + b$. 
If the synthetic spectrum is the same as the observations, $\rho = b =1$ and $\alpha = 0$. 
$\rho$ is a measure of magnitude comparison, and we regard empirically
that the synthetic and observed spectra agree with each other when $\rho$ ranges between 0.7 and 1.3. 
$\alpha$ is an indication of the plasma temperature distribution, since the thermal bremsstrahlung spectrum
is temperature dependent. When $\alpha > 0$, the model has produced more high temperature plasma than 
observed, and when $\alpha < 0$, there is a lack of hot plasma in the modeled DEM.

In these plots, we only compare the spectrum up to 20 keV, beyond which the spectrum is likely dominated by 
thick-target bremsstrahlung produced by non-thermal electrons colliding at the lower atmosphere.
For most of the impulsive phase, from 16:44~UT to 16:59~UT, the synthetic spectra 
agree with the observations (see panels b, c, and d in Figure~\ref{Fig10}), with mean 
$\rho$ value between 0.7 and 1.3 and $\alpha$ close to zero. From 16:50 to 16:55~UT, around the maxima
of the {\it GOES} SXR light curves, the modeled thermal emission dominates up to 20~keV (see panels c and d), indicating
strong heating in flare loops although the non-thermal HXR emission has become insignificant.
These plots showing good agreement between the model and observations suggest that the model 
has rather accurately reproduced the DEM responsible for the thermal bremsstrahlung X-ray
radiation in the observed range during this period. 

From the start of the flare until 16:42~UT, the synthetic spectrum is 
lower than observed by nearly the same fraction across the energy range up to 15~keV (panel a in Figure~\ref{Fig10}). 
The synthetic flux at $>$15~keV is still smaller than observed, which may indicate a
lack of high-temperature plasma in the model, but may also be due to predominant non-thermal bremsstrahlung 
radiation down to 15~keV. {\it RHESSI} images in the 15--20~keV energy range during this period (not shown) show that
the X-ray source is rather extended, likely including emissions from both the foot-point and loop-top, which
cannot be resolved with good photometric accuracy. Therefore, both scenarios are plausible.
In the late phase of the flare after 17:00~UT, the modeled flux drops below the observed
flux again. In this phase, thick-target HXR emission has ended, so the low flux above 15~keV is likely caused by a
lack of high temperature plasma in the model. We also note that in the early and late phases of the flare, when the average 
ratio $\rho$ is far below 1, the slope $\alpha$ becomes negative. These indicate that the model does not generate 
enough high temperature plasma during these periods, maybe due to a lack of heating events identified in our method.
The bottom panel of the figure shows the number of heating events identified during the flare. It is
seen that the number of heating events drops rather significantly both before 16:39~UT and after 16:55~UT. It appears
that the shortage of heating events leads to insufficient X-ray flux in the calculation about 3 to 5 minutes later, which
is roughly the timescale for temperature and density increases in newly formed flux tubes. The lack of heating events 
identified from lower atmospheric signatures (brightened UV or HXR foot-points) could be explained by the fact that 
there is neither strong conduction flux nor significant non-thermal beam to deposit enough energy 
at the lower atmosphere to generate observable signatures. In the rise phase, it is plausible despite the significant 
non-thermal HXR emission, if the non-thermal electrons are partially or largely trapped in the corona and do not reach 
the chromosphere to generate HXR foot points; the trapped electrons would then be a source of {\it in situ} heating 
in the corona.  Such a mechanism has been proposed for the ``pre-impulsive'' phase of a few large flares 
\citep[e.g.,][]{Lin03,CaspiPhDT,Caspi10} where non-thermal emission is observed in the corona with no identifiable HXR foot points, 
and may also contribute here.

Comparison between the observed and synthetic spectra, as characterized by $\rho$ and $\alpha$, provides us with
observational constraints to the heating function parameters $\lambda$ and $\gamma_m$, in addition 
to the observed UV and HXR signatures at the flare foot-points that constrain the time profiles
of the heating rate. 

\subsection{UV Emission in the Decay Phase}\label{Sec4.3}
The long decay of {\it TRACE} 1600~\AA{} emission in individual flaring pixels is reported 
by \citet{Qiu10} and \citet{Cheng12}. Flare emission in this broadband channel includes both the enhanced UV continuum and the C~{\sc iv} 
line emission. The C~{\sc iv} resonance doublet at 1548~\AA{} and 1551~\AA{} could only be emitted over a very narrow temperature 
range around 10$^5$~K, and therefore the C~{\sc iv} 
intensities rise when the transition region is heated with more plasma raised to 10$^5$~K
during the flare. The C~{\sc iv} irradiance was observed to be enhanced by three to four orders 
of magnitude over the pre-flare emission during the impulsive phase of a large flare \citep{Brekke96}. 
During the decay phase, the C~{\sc iv} emission is dominated by the evolution of coronal pressure 
\citep{Fisher87, Hawley92, Griffiths98}. The solar UV continuum radiation in the 1000--2000~\AA{} wavelength range is also enhanced during the flare \citep[e.g.,\ ][]{Cook79,Cheng84,Cheng88}. In particular, the continuum emission below 1682~\AA{} is primarily contributed
by bound-free transitions of Si~{\sc i}, which is primarily excited by the C~{\sc iv} doublet \citep{Machado82, Machado86}. 
Therefore, the continuum intensity at $\lambda <$ 1682~\AA{} is approximately proportional to  the C~{\sc iv} line intensity \citep{Phillips92}. 
Though there is no direct C~{\sc iv} measurement for the M8.0 flare studied in this paper, observations of stellar 
flares show that the time profile of C~{\sc iv} is similar to the UV 1600~\AA\ emission in solar flares observed by {\it TRACE}, both exhibiting 
a fast rise and a long decay \citep{Vilhu98}. 

In this study, we have used the rapid rise of UV emission to constrain the time, duration, and magnitude of the heating rate
in individual flare loops. The gradual decay of UV emission from the same foot-points reflect the evolution of the over-lying
corona that has been heated impulsively and then cools down on a much longer timescale. Using the coronal plasma properties
from the model, we further calculate the C~{\sc iv} line emission during the decay phase to compare with the {\it TRACE} UV observation.

To calculate the C~{\sc iv} line emission on the solar surface, it is important to know the opacity. 
\citet{Doschek91} and \citet{Dere93} measured the intensity ratio of two C~{\sc iv} lines  at $\lambda$1548 and $\lambda$1550
observed on the disk, and found the ratio very close to 2:1, which is expected for optically thin lines. 
\citet{Doschek97} also estimated the opacity of C~{\sc iv} to be 0.099 in the active-region spectrum. 
We therefore consider that the C~{\sc iv} line is optically thin. The C~{\sc iv} photon flux at the surface is then calculated through 
\begin{equation}
I_{\text{C~{\sc iv}}} = A_b\int _0^\infty C(T, n) \textrm{ DEM}(T)\; \mathrm{d} T \text{~photons~cm$^{-2}$~s$^{-1}$~sr$^{-1}$}\label{eq_civ}
\end{equation}
where $A_b$ is the abundance of Carbon relative to Hydrogen. In this paper the solar coronal abundances of \citet{Feldman92} 
are used; if photospheric abundances \citep[i.e.,][]{Grevesse98} are used, the C~{\sc iv} line emission will be reduced by 15\%.
$C(T, n)$ is the contribution function for the C~{\sc iv} line and is obtained from CHIANTI \citep{Dere97, Landi12} 
with the ionization equilibrium of \citet{Mazzotta98}. It does not change significantly when  the electron density $n$ varies 
from $10^9$~cm$^{-3}$ to $10^{12}$~cm$^{-3}$. DEM$(T) \equiv n^2\left(\frac{\partial T}{\partial s}\right)^{-1}$ is the transition region DEM, 
where $s$ measures distance along the flux tube from the base of the transition region. 
For each flux tube, we assume that the transition region is nearly in static equilibrium in the decay phase when the 
heating has finished, which is roughly when the coronal density peaks. In this case, thermal conduction is balanced by radiation, 
and pressure is uniform from the corona through the transition region when gravity is ignored, and the transition 
region DEM can be calculated as \citep{Fisher87,Hawley92}:
\begin{equation}
\textrm{DEM}(T) = P \sqrt{\frac{\kappa_0}{8k_{\rm B}^2}} T^{1/2}\zeta^{-1/2}(T) \label{eq_dem}
\end{equation}
where $\zeta(T) = \int _{T_0}^T T'^{1/2} \Lambda(T')\; \mathrm{d} T'$, $T_0$ is the temperature at the base of the transition region, taken
to be a nominal 10$^4$~K, and $\Lambda(T)$ is the optically-thin radiative loss function. 

\begin{figure}
 \begin{center}$
 \begin{array}{cc}
 \includegraphics[width=3.in]{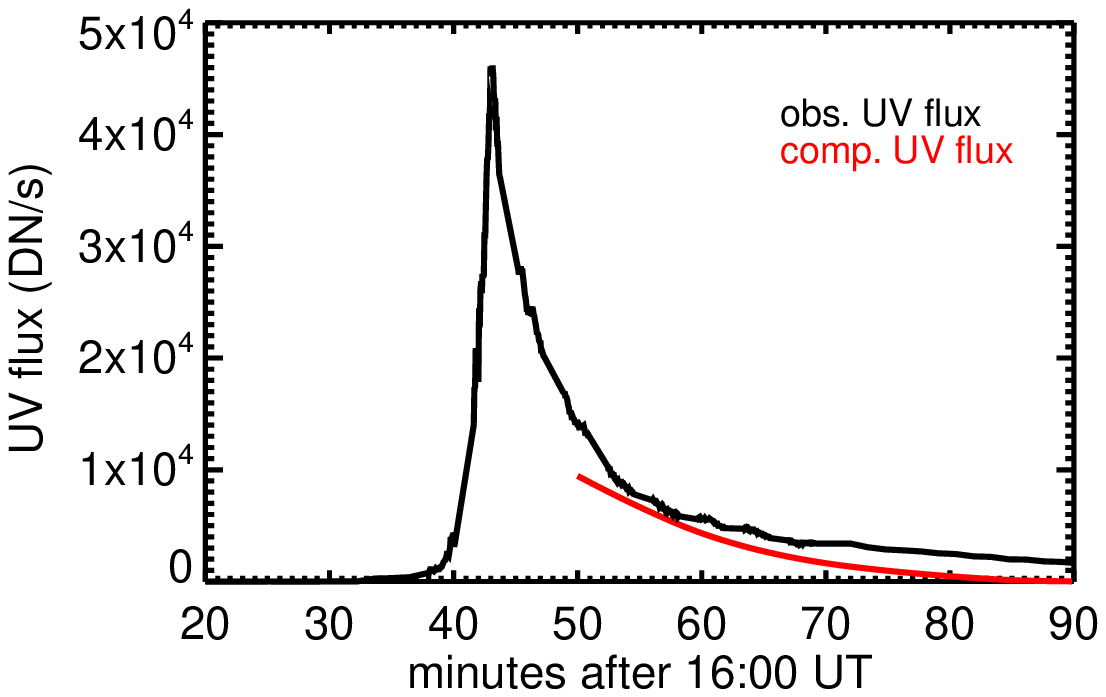} &
 \includegraphics[width=3.in]{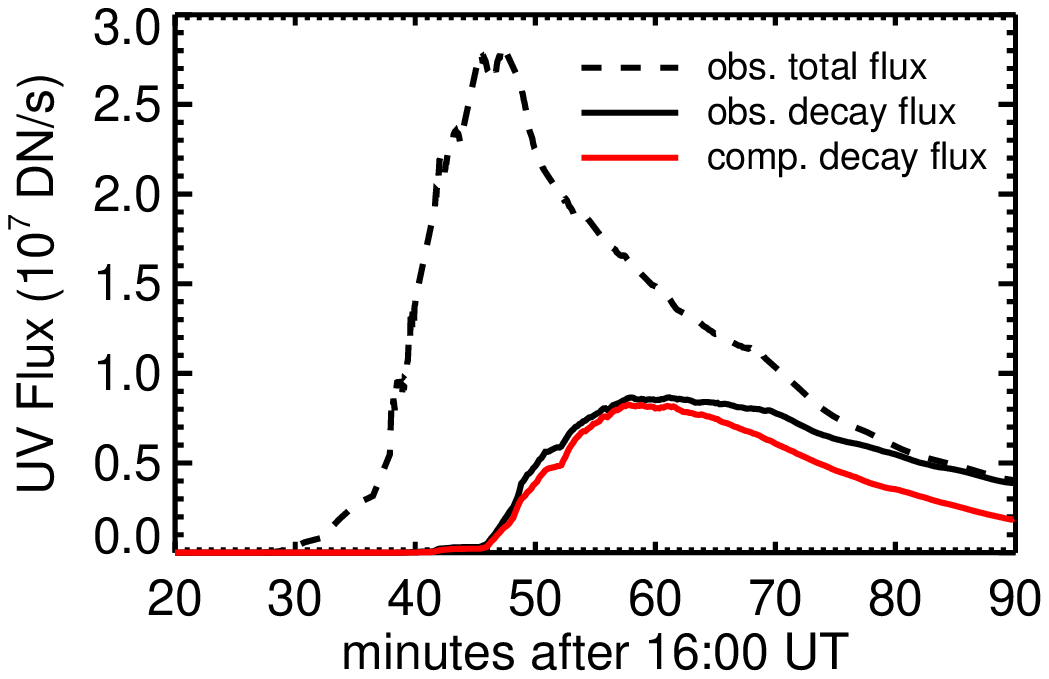}
 \end{array}$
 \end{center}
\caption{Comparison of the synthetic transition region C~{\sc iv} line emissions (red) 
in the decay phase with the UV flux observed in the {\it TRACE} 1600~\AA{} band (black) 
for one flare loop (left) and for all flare loops (right). In the right panel, the observed 
decay flux is derived by subtracting from the observed light curve the full Gaussian profile 
that represents the UV rise. } \label{Fig12}
\end{figure}

The C~{\sc iv} line emission of one flux tube in its decay phase is calculated with equations (\ref{eq_civ}) and (\ref{eq_dem}) 
and convolved with the {\it TRACE} 1600~\AA{} band response function to get synthetic UV flux in units of DN~s$^{-1}$.
The left panel of Figure~\ref{Fig12} shows an example of the computed C~{\sc iv} line emission in one flux tube. The computed C~{\sc iv} flux
appears to evolve along with the observed decay, though the synthetic flux is smaller than the observed by a factor of 2. 
The right panel of the figure shows the sum of the synthetic UV flux in all flaring pixels in comparison with the observed total UV emission 
in the decay phase. The observed UV decay in a pixel is obtained by subtracting from the UV light curve 
the full Gaussian derived by fitting the UV rise so that the transition region response to the heating is excluded. 
Similar to the single-pixel comparison, the computed total C~{\sc iv} emission decays
on the same timescale as observed, and the computed flux is lower than observed by a factor of 2 to 3. 

We find this comparison satisfactory for the following reasons. First, the transition region is not exactly in static equilibrium. 
During the decay phase, the transition region loss usually exceeds the conductive flux, which leads to downflows from the corona, 
and therefore the coronal density gradually decreases. The downflow is literally an energy input into the transition region and 
gives rise to a greater transition region DEM than in the case of static equilibrium. Second, the UV continuum in the decay phase 
may be also enhanced due to irradiation from the C~{\sc iv} line. The observed UV emission is the sum of both the line and
continuum emissions, though the continuum intensity is considered to be proportional to the line intensity. 
All these reasons would explain the observed UV emission being larger than the computed C~{\sc iv} line emission with the static equilibrium
assumption. It is, nevertheless, very encouraging to find that the computed line emission decays at the same rate as 
the observed UV emission. In contrast, the computed X-ray emission decays faster than observed, as shown in the previous
Section. Since the UV decay is derived from the same pixels whose rise phase is used to construct the loop heating rate, 
such comparison indirectly confirms that the rapid decay of the computed SXR flux most likely indicates the presence
of additional heating events in the decay phase, which are not found in foot-point UV radiation signatures.
This is consistent with decades of observations which have shown that coronal plasma temperatures decrease more slowly than would be expected
from simple calculations of radiative and conductive cooling \citep[e.g.,][]{Moore80,Veronig02a,Veronig02b}, implying that additional heating must
be taking place despite an apparent lack of accelerated particles.
These additional heating events may thus result from continuous reconnection in the corona that provides direct heating without
significantly accelerating non-thermal particles, or which accelerates particles sufficiently weakly such that they thermalize in the ambient plasma at the loop-top; both scenarios lead to increased SXR coronal emission without associated chromospheric emission that
could be identified in {\it TRACE} broadband UV or {\it RHESSI} HXR observations.  A similar mechanism has been proposed for the so-called ``EUV late phase'' emission from eruptive (and often CME-associated) flares \citep{Woods11}, where
reconnection high in the corona heats post-flare loops, resulting in increased EUV coronal emission without associated foot-point brightenings.

\section{Discussion and Conclusions}\label{Sec5}

We have analyzed and modeled an M8.0 flare on 2005 May 13 observed by {\it TRACE}, {\it RHESSI}, and {\it GOES}, in order
to determine heating rates in a few thousand flare loops formed by reconnection and  subsequently heated.
The rapid rise of spatially resolved UV emission from the lower atmosphere is considered to be the 
instantaneous response to heating in the flare loops rooted at the foot points, and therefore is used to construct heating rates 
in individual flare loops to describe when, for how long, and by how much flare loops are heated \citep{Qiu12}. 
With these heating rates, we compute the coronal plasma evolution in these loops using the 0D EBTEL model, 
and compare, for the first time, the computed time-dependent coronal radiation with observations by {\it RHESSI} and {\it GOES},
and as well the computed UV line emission with {\it TRACE} observations during the decay phase of flare loops.
The comparison constrains parameters of the loop heating rates, which define the amount 
of total energy flux and the fraction of energy flux carried by chromospheric upflows driven by non-thermal beams.

The M8.0 flare studied in this paper is observed to have significant thick-target HXR emission, 
suggesting that the lower atmosphere is also heated by electron beams during the impulsive phase. 
In this process, chromospheric evaporation is driven carrying mass and energy flux back to the corona. 
In this study, we experimentally scale the beam-driven energy flux with the observed HXR light curve, and 
examine how varying the fraction of beam heating changes the modeled coronal temperature and density evolution. 
In general, for a given total amount of heating flux, a larger fraction of energy carried by beam 
driven upflow leads to higher density, lower temperature, and more rapid evolution of the coronal plasma.
These properties affect the computed SXR spectrum and its time evolution, and can be compared against the spectrum
observed by {\it RHESSI}. Our experiment yields an optimal set of parameters, with which, the model computed SXR light 
curves and spectra agree very well with observations during the rise phase and peak of the flare for more than 10 minutes.

Apart from comparison with the SXR observations, we also compute the C~{\sc iv} line emission
at the foot of the flare loop during the decay, whose rise phase is used to construct the 
heating rate in the same loop. Emission in this optically-thin line during the decay phase is governed by 
plasma evolution in the overlying coronal loop and contributes significantly to the UV emission observed
in the {\it TRACE} 1600~\AA\ bandpass. It is shown that the computed C~{\sc iv} flux is about one third of
the observed UV flux, which also includes the continuum radiation, and decays on the same timescale
as observed. This experiment shows an avenue to model heating and cooling
of spatially resolved flare loops with self-consistent constraints by high-resolution UV 
observations from the input (UV rise) to the output (UV decay) in the same flare loop.

These results suggest that our method, which employs all available observations
to constrain flare loop heating model, is able to capture the distribution of impulsive 
heating rates in numerous flare loops formed and heated sequentially by magnetic reconnection. 
The optimal heating function parameters determined for this flare yield the peak heating flux
in the range of 10$^{8}$--10$^{10}$~ergs~cm$^{-2}$~s$^{-1}$ for over 5000 flare loops of cross-sectional
area 1\arcsec\ by 1\arcsec. Notably, only $\lesssim$40\% of this energy is carried by beam-driven upflows
during the impulsive phase, with the remaining $\gtrsim$60\% contributed by {\em in situ} heating 
in the corona. The beam heating occurs mostly during the first ten minutes, 
when the flare exhibits significant thick-target HXR emission; meanwhile, the {\it in situ} 
heating energy in the impulsive phase amounts to 6.2
$\times$10$^{30}$~ergs, over 3 times greater than the beam-driven contribution. 
In the post-impulsive phase (after 16:50~UT), the continuously expanding UV ribbons and the observed Neupert effect 
between the SXR derivative and UV (rather than HXR) light curves indicate that new flare loops are formed by continuous 
reconnection, and {\it in situ} heating in the corona is predominant, amounting to 4.1
$\times$10$^{30}$~ergs out of the total, with negligible beam heating. The total energy used in coronal heating in this flare amounts to 1.22
$\times$10$^{31}$~ergs, of which the total energy carried by beam-driven upflows is estimated to be only 1.9
$\times$10$^{30}$~ergs ($\sim$16\%). Therefore, it appears that {\it in situ} coronal heating and 
thermal conduction play an important role in the energetics of this M8.0 flare, in contrast to 
commonly accepted models that consider the coronal thermal plasma to be largely a by-product of 
beam-driven heating of the lower atmosphere (chromosphere and transition region)
\citep[see also discussions by][]{Longcope10, Caspi10, Longcope11}.
A survey of 37 M- and X-class flares \citep{Caspi13} suggests that {\it in situ} heating 
may be significant even down to mid-C class, and therefore this heating mechanism must be considered 
in future studies of flare energetics.

We note that these energies set lower limits for this M8.0 flare. First, despite
the excellent agreement between the model and observation in the rise phase of the flare, 
the model computed X-ray flux decays more quickly than observed by both {\it GOES} and {\it RHESSI}. It is most likely
that magnetic reconnection and energy release continue in the high corona during the decay of the flare.
These heating events might not produce significant UV foot-point emission, and are therefore not identified
in our method. For comparison, \citet{Kazachenko09,Kazachenko12} analyzed the same flare, and estimated the 
total energy from {\it GOES} (including energy loss from radiation, conductive cooling and enthalpy flux) to be 3.1$\times$10$^{31}$ ergs.  
Second, in our experiment, we do not model heating and dynamics of the lower atmosphere, but use a simple scaling
relationship to estimate the energy flux carried by non-thermal beam-driven upflows back to the corona. This
amount of energy is a fraction of the total energy carried by non-thermal beams; the remainder is 
lost in the lower atmosphere. As a reference, we fit the HXR spectrum to a power-law distribution 
with the standard {\it RHESSI} software package (not shown in the paper), and, with the low-energy cutoff and 
spectral index from the fit, estimate the total non-thermal energy to be 7.6$\times 10^{30}$~ergs, about
4 times the energy carried in the upflow but, interestingly, still somewhat smaller than the energy required to heat the coronal plasma --
though we note that the non-thermal energy estimate is only a loose lower bound since it depends critically on the fit
low-energy cutoff, which is unbounded from below due to obscuration by the highly-dominant thermal emission.

Our experiment provides a novel method for investigating 
energy release in solar flares, which is governed by 
reconnection and substantiated in formation and heating of numerous flare loops. 
This approach has a few advantages. First, by analyzing the foot-point UV signatures combined with HXR observations, 
for the first time, we are able to identify and characterize the energy release process in a few thousand flare loops down to 
1\arcsec\ by 1\arcsec\ scale, which is thought to be close to the basic scale of flux tubes formed in 
patchy reconnection. Using the {\em same} UV data, we also measure the rate of magnetic reconnection and
hence are able to establish the relationship between reconnection and energy release in temporally and 
spatially resolved manner. Second, using the 0D EBTEL model, we are able to efficiently compute plasma
properties of this large number of flux tubes, and naturally generate a time-dependent differential 
emission measure (DEM) from these flare loops formed and heated at different times and evolving independent of one another. 
Subsequently, we can predict X-ray and UV radiation signatures directly comparable with observations, which 
provides further constraints to our determination of energy release rates. 

The method utilizes emission signatures at all available wavelengths in both the foot-points and overlying coronal loops. 
The dynamics of the lower atmosphere and corona are strongly coupled; they are governed by different yet coherent physics, 
and hence should be studied coherently. In this study, based on the insight from
previous theoretical and numerical research, we employ some empirical laws to prescribe the 
relation between heating rates and radiation signatures. The optimal parameters obtained from the
analysis can provide a reference for further investigation based on physical models, such
as models of lower atmosphere heating and dynamics. We also recognize that the 0D EBTEL model
has limitations in accurately describing coronal plasma properties along the loop. Nevertheless, the
model is highly efficient in dealing with a few thousand flare loops and yields the first-order heating rates as
useful inputs for more sophisticated loop heating models such as those by \citet{Mariska87}, \citet{Antiochos99}, 
or \citet{Winter11}. Our study of this event shows very good agreement between observed and modeled high 
temperature plasma radiations (e.g., by {\it RHESSI} and {\it GOES}), but lacks observations of and therefore 
comparison with low temperature plasma signatures, which are crucial in understanding the flare loop evolution 
beyond the conduction-dominant regime. The recently launched {\it SDO} has observed many flares 
with AIA, the high-resolution imaging telescope, in a number of UV and EUV bands \citep{Lemen12}, 
and as well with the  broad-band, high resolution spectrometers on the Extreme Ultraviolet 
Variability Experiment \citep[EVE;][]{Woods12}. These new observations, which cover coronal temperatures from $\lesssim$1~MK up to $\sim$20~MK, can be combined 
with RHESSI observations that provide information of hot ($\gtrsim$10--50~MK) plasmas, as well as information 
about the non-thermal emission, to fully observationally characterize the temperature distribution and its evolution (Caspi et al.~2013b, in prep.), which can be compared with -- and used to constrain -- our modeling. These new facilities will allow us to 
analyze and model flare plasmas following the entire process of heating and cooling, and also to study 
resolved individual flare loops in multiple wavelengths in more details, which will improve determination
of the energy release rate during magnetic reconnection. 

Finally, the poor comparison between model and observations during the decay phase of the flare
reflects some limitations of the present method. These can be addressed by more sophisticated one-dimensional 
modeling that also includes the lower atmosphere, to help determine the amount of lower atmosphere radiation, 
such as the UV 1600 {\AA} emission used in this study, as dependent on the heating mechanism and the amount of heating.
Such effort will help clarify whether the shortage of computed flux is produced by the method
missing weak heating events, or by other significant effects related to imperfect model assumptions.

\acknowledgments We thank Drs. J. Klimchuk and G. Fisher for insightful discussion and the anonymous
referee for thoughtful and valuable comments.
The work is supported by NSF grant ATM-0748428. 
A. Caspi was supported by NASA contracts NAS5-98033 and NAS5-02140, and NASA grants NNX08AJ18G and NNX12AH48G.
CHIANTI is a collaborative project 
involving the University of Cambridge (UK), George Mason University, 
and the University of Michigan (USA).


\end{document}